\documentclass[3p]{elsarticle}

\usepackage{amssymb}
\usepackage{amsmath}
\usepackage{graphicx}
\usepackage{ucs}
\usepackage{url}
\usepackage[utf8x]{inputenc}
\usepackage{subfigure}
\usepackage{booktabs}

\DeclareMathOperator*{\argmin}{argmin}
\journal{Physica A}

\begin{document}

\begin{frontmatter}



\title{A maximum entropy network reconstruction of macroeconomic
models}

\author[label1,label2]{Aurélien Hazan}
\address[label1]{Université Paris-Est}
\address[label2]{LISSI, UPEC, 94400 Vitry sur Seine, France}



\begin{abstract}
In this article the problem of reconstructing the pattern of connection 
between agents from partial empirical data in a macro-economic 
model is addressed, given a set of behavioral equations.  

This systemic point of view puts the focus on distributional and network effects,
rather than time-dependence. Using the theory of complex networks we 
compare several models to reconstruct both the topology and the flows of money
of the different types of monetary transactions, while imposing a series
of constraints related to national accounts, and to empirical network sparsity.
Some properties of reconstructed networks are compared with their empirical
counterpart.
\end{abstract}


\end{frontmatter}


The state and dynamics of production and financial sectors are affected
by interconnection patterns between agents. However these patterns are 
poorly known at a detailed level because of confidentiality issues. This
prevents building risk indicators, that can help preventing the
propagation of crises. 

Our main motivation here is to infer an ensemble of unobserved multilayer
networks in a data-scarce context, taking advantage of available
information: first, aggregate public statistics are used to estimate the
number and importance of the nodes involved. Then, macro-economic models
established by economists provide consistency and behavioral constraints
that further reduce the size of the reconstructed ensemble.

A second motivation is to recover empirical regularities from the
reconstructed ensemble, that are also available in a few empirical
studies, which will help assessing our method. For example, the degrees
of firms in buyer-supplier networks have been the subject of many studies,
and are hard to measure and to model. They are fundamental properties of
the network that influence the dynamical processes that take place on it.
The same holds for strengths as functions of degrees.
Even though the macro-economic model used below is theoretical, it allows
to compute these empirical regularities from the reconstructed ensemble.
This  paves the way for more detailed models with more sectors
(government, central bank, financial services, the environment, etc...),
and more refined mechanisms.

A third motivation, not developed in this article and left for future
works,  will be to explore the consequences of building such a
reconstructed ensemble in terms of economic applications:  define and
estimate risk levels in production networks, study the risk of crisis
propagation between the productive and financial sectors, identify hidden
clusters among nodes, etc\ldots

To do so our method in this paper builds on advances in the field of
data-driven network reconstruction from partial information. While
actively investigated in recent years, it remains an open problem. 
Maximum entropy methods were introduced in statistical physics
\cite{jaynes_information_1957} and were shown to be very useful in 
studying the properties of network ensembles \cite{park_statistical_2004}.
Their application to social systems 
\cite{hernando_maxent_2012,hernando_variational_2012, hernando_workings_2013,zambrano_thermodynamics_2015}, to financial and economic networks is well established \cite{squartini_maximum-entropy_2017}
and was recently demonstrated by central bankers to be rather
accurate \cite{anand_missing_2018}.

Network theory which deals with the structure and dynamics of network,
and the properties of dynamics over networks has been prolific overs past
years \cite{newman_networks:_2010} and seen a rising interest among
economists \cite{jackson_social_2013}. Theoretical models of supply
chain, credit \cite{battiston_credit_2007}, trade, as well as empirical
studies have covered various topics such as employment, world trade or
ownership control among corporations \cite{vitali_network_2011}.

However, up to our knowledge, the possibility to include constraints 
inspired by macroeconomic models in Maxent network models was little
studied so far. 
In engineering for example, network reconstruction based on physical
laws was developed for flow networks \cite{waldrip_comparison_2017}.
In the present article we propose to introduce such a constraint on the
network ensemble, with an unknown topology.

The most direct way to reach that aim would be to extend one of the
existing weighted reconstruction methods such as 
\cite{mastrandrea_enhanced_2014, cimini_estimating_2015}, but this
extension raises several issues.
Rather, we use a two-step method that requires first to estimate the
topology of each subnetwork taken independently, from partial empirical
data, in the spirit of \cite{garlaschelli_fitness-dependent_2004}.
This choice stems from the fact that weighted networks estimation is 
greatly improved when topological information is available.
Then, when the topology is reconstructed, network weights can be
approximated using various methods, either probabilistic or deterministic.

The highlights of our method to reconstruct an ensemble of networks follow:
i) it is probabilistic, ii) relies on Maximum Entropy, iii) enforces
non-negativity of weights, iv) allows linear constraints on connections
and weights, of the form  $A\xi-b=0$. More specifically :
\begin{itemize}
\item it is probabilistic because deterministic methods do not allow
sampling random instances, that are necessary to compute average
quantities.
\item Maximum Entropy makes it possible to respect empirical measurement
(e.g. aggregate weights, link density), while maximizing randomness,
to avoid biases.
\item non-negativity of weights is required by the nature of economic
transactions.
\item linear constraints in the form $A\xi-b=0$ incorporates consistency
constraints that stems from models built by economists.  
\end{itemize}
Lastly we stress that when detailed micro data is available, a 
reconstruction error can be computed to assess the accuracy of the method.

In section \ref{sec.sfc.model} we present the testbed model for this study,
which is inspired by stock-flow consistent (SFC) macroeconomic models, in
the restricted linear and steady-state case. In section \ref{sec.network.reconstruct}
the corresponding binary network is defined and reconstructed. 
In section \ref{sec.marginal.pdf}, the weights are computed, using 
several methods. Sections \ref{sec:comparison} and \ref{sec.discussion}
compare and discuss the results while section \ref{sec.conclusion} concludes.




\section{A disaggregated linear model}
\label{sec.sfc.model}

In this section we describe a toy macroeconomic model, only consisting
in a transaction matrix that defines the flows of money between
origins and destinations, as shown in Tab. \ref{sfc.transactions}.
There are no stocks (capital, loans, deposits) nor balance sheet,
nor behavioral equations responsible for consumption, capital depreciation.
It is a simplification of the aggregated BMW model proposed in
\cite[§7]{godley_monetary_2007}, an SFC model that introduces
private bank money and does not involve a state nor a central bank, and 
where each sector is represented by a single agent.

However, our model deals with the disaggregated case: the number of 
households, firms and banks is arbitrary. Households can buy from several
firms, get wages from various employers and interest on deposits from 
several banks. Firms can buy capital goods from many firms, and pay interest
on loans to several banks.

\begin{table}[htbp]
\centering
	\begin{tabular}{llllllll} %
	\hline		
 	& \multicolumn{3}{c}{Households}  & \multicolumn{2}{c}{Firms} & Banks & $\sum$ \\
 	\cmidrule(lr{.75em}){2-4}  \cmidrule(lr{.75em}){5-6} \cmidrule(lr{.75em}){7-7}
	  & 1 & 2 & 3 & 1 & 2 &1 & \\
	\hline
	Consumption & -$C_{d11}$ &-$C_{d21}$  &  & $C_{s1}$ &   &    & 0\\
	            &  & -$C_{d22}$ &  & $C_{s2}$ &   &    & 0\\
	            & -$C_{d13}$ &  & -$C_{d33}$ & $C_{s3}$ &   &    & 0\\
	\hline            
	Investment &  &  &  &  -$I_{d1}$ & $I_{s1}$ &    & 0\\            
	\hline            	
	Wage & $WB_{s1}$ &  &  & -$WB_{d1}$   &  &    & 0\\
	     &  & $WB_{s2}$ &  & -$WB_{d2}$   &  &    & 0\\
	     &  &  & $WB_{s3}$ &   & -$WB_{d3}$  &    & 0\\
	\hline            	
	Interest on loans &  &  &  & -$IL_1$ & &  $IL_1$  &  0\\            
	\hline            	
	Interest on deposits & $ID_1$ &  &   &  & &  -$ID_1$   & 0\\            
					     &  & $ID_2$ &   &&  & -$ID_2$   &   0\\            
						 &  &  & $ID_3$   &  &  & -$ID_3$    & 0\\            	
	\hline
	$\sum$ & 0 & 0 & 0 & 0 & 0 & 0   & 0\\            
	\hline
	\end{tabular}
\caption{\label{sfc.transactions}Transaction matrix of a disaggregated model 
with many households, two firms and one bank: agents $nh=3$, $nf=2$, $nb=1$. 
See Tab. \ref{tab.bmw short labels} for notations.
}
\end{table}

To simplify further, and keep the focus on topological and distributional 
effects, the systems is supposed to be in the steady-state\footnote{a
detailed study in \cite{godley_monetary_2007} establishes the properties
of the BMW model in the transient and steady-state regimes,
in the case of the representative model that is when each institutional 
sector is represented by one agent.} regime.
The set of row and column-sums equations in Tab. \ref{sfc.transactions}
can be written in the form a linear system: 
\begin{equation}
S=\{ \xi ~s.t. ~A \xi=b  \}
\label{eq.syslin}
\end{equation}
Furthermore the system can be written as a function of demands only: 
\begin{equation}           	
\xi=
\begin{bmatrix}
Cd&Id&WBd&ILd&IDd 
\end{bmatrix}^T
\end{equation}
with vectors $C,I,WB,IL,ID$ standing for consumption, investment, wage bill,
interest on loans, interest on deposits and subscripts $d$ and $s$ referring to 
demand and supply as summarized in Tab. \ref{tab.bmw short labels}.
The vectors are indexed so as to encode all origin-destination information:
\begin{equation}           	
Cd=
\begin{bmatrix}
Cd_{1,1} & \ldots &Cd_{n_h,1} & Cd_{1,2} & \ldots &Cd_{n_h,2} &  \ldots& Cd_{1,n_f} & \ldots &Cd_{n_h,n_f} 
\end{bmatrix} 
\end{equation}
where $Cd_{n_h,n_f}$ is the amount of consumption goods sold by firm $n_f$ 
to household $n_h$.

The same indexing convention is adopted for $I,WB,IL,ID$, as explained in
\ref{appendix.notations.syslin}. This leads to the following expression for $A$
and $b$:
\begin{equation}  
A=
\begin{bmatrix}
-I_{h1}& & I_{h2} && I_{h3}\\ 
I_{f1}&I_{f2} & -I_{f3} &-I_{f4}& \\ 
& & &I_{b1} & -I_{b2} \\ 
\end{bmatrix},
~b=0_{n_b+n_f+n_h}
\end{equation}

The submatrices $I_{*}$ reflect the connection pattern among agents, in
a way such that each line of $A$ can enforce the constraints in Tab.
\ref{sfc.transactions}. 
$A$ is a matrix with elements having
values in $\{-1,0,+1\}$, 
composed of $2 n_f n_h + n_f^2 + n_b(n_f+n_h)$ columns and
$n_b+n_f+n_w$ rows. Its sparsity factor is close to $10^{-3}$ given the
values of the parameters. The system $A\xi=b$ is under-determined.

Note that the system eq.(\ref{eq.syslin}) is homogeneous and has a 
trivial solution. For this reason, it will be useful below to define
a minimalist nonhomogeneous system: 
\begin{equation}
S_1=\{ \xi ~s.t. ~A_1 \xi=b_1  \}
\label{eq.syslin.alpha0}
\end{equation}
where $A_1$ and $b_1$ are such that the consumption of all 
households is set to the constant value $\alpha_0$.

How the connection pattern is reconstructed from partial
empirical data is the subject of sec. \ref{sec.network.reconstruct}.

\begin{table}[htbp]
\centering
\begin{tabular}{ll}
\hline
 Variable & Label\\
\hline
 loans to firms  & $L$   \\
 investment & $I$  \\
 interest on loans & $IL$  \\
 wage bill & $WB$   \\
 interest on workers deposits & $ID$  \\ 
 consumption of workers & $C$  \\
\end{tabular}
\caption{Labels associated with the different monetary variables, after 
\cite{godley_monetary_2007}. The subscripts $d$ and $s$ stand for demand and
supply.}
\label{tab.bmw short labels}
\end{table}


\section{Random network reconstruction and sampling}
\label{sec.network.reconstruct}

In order to parametrize the disaggregated model in sec. \ref{sec.sfc.model},
empirical datasets are necessary. Detailed datasets exist in some
particular cases: consumption networks were studied in Latin America
and Europe \cite{dong_social_2017,leo_correlations_2018}.
The buyer/supplier interfirm network in Japan
\cite{watanabe_economics_2015,mizuno_buyer-supplier_2015}, in 
Estonia \cite{rendon_de_la_torre_topologic_2016}, in the 
USA \cite{atalay_network_2011} where the distribution of supply chains
was modelled by a birth-death process.
The ownership network of transnational companies was reconstructed in \cite{vitali_network_2011} 
whith spatial distances as an explanatory variable \cite{vitali_geography_2011}.
Apart from interfirm links, a study of Japanese bank-firms relationships can be 
found for example in \cite{watanabe_new_2015}.

However, when detailed transaction databases exist, their access is
restricted or paywalled. Only aggregated ones are publicly available
for most developed countries. 

The topic of input/output relations between industrial sectors
has been studied by economists dating back to Leontief \cite{w._leontief_quantitative_1936}.
At the aggregate level, this weighted network is densely connected,
as remarked in \cite{blochl_vertex_2011}, but dense connection is not 
a property observed at the micro level in available detailed datasets. 
The same kind of problem arises for other types of networks, which reveals 
the necessity to perform network reconstruction \cite{squartini_maximum-entropy_2017}.

Before describing the reconstruction methods used below, let us define 
some notations: a binary graph $G$, with at most one
edge $e \in E$ between two vertices in $\mathcal{I}\times\mathcal{J}$, is 
specified by its 
adjacency matrix $\mathbf{A}=\{ a_{ij} \}_{i \in \mathcal{I}, j \in \mathcal{J}}$. 
This covers the unipartite case when $\mathcal{I}=\mathcal{J}$.
The degree sequences will be denoted by $k$. For graph ensembles,
$p_{ij}$ is the connection
probability between vertices $i$ and $j$, and $\langle . \rangle$ denotes
an average over that ensemble.
The n-uple of adjacency matrices that correpond to the model in 
sec. \ref{sec.sfc.model} is $\mathcal{A}=(\mathbf{A}_{cons},\ldots,\mathbf{A}_{invest})$.
Lagrange multipliers in maximum-entropy models will be written in the form
$x_i$, with $i \in \mathcal{I}$.

Popular graph examples include the Erd\"os-Rényi random
graph model (generalized to the bipartite case under the name BiRG, for
Bipartite Random Graph \cite{straka_grand_2017}), and the configuration model (CM, see \cite[2.2.2]{squartini_maximum-entropy_2017}). 
The latter defines an maximally random ensemble of graphs such that the
degree $k_i$ of each vertex is constrained to experimental values, 
that is in the undirected case: 
$\forall i \in \mathcal{I}, \langle k_i \rangle=k_i$.
However such detailed information is not available in our case.
To go round this difficulty the notion of fitness $g_i$ was
put forward in the litterature \cite[p.82]{squartini_maximum-entropy_2017}. 
The hypothesis made is that ``the probability for any two nodes to interact
can be explicitly written in terms of non-structural quantities'',
which suggests to write the Lagrange multiplier in the form $x_i = f(g_i)$,
where $g_i$ are ``non-structural quantities'' such as the trade volume between
countries.  

Node-specific fitnesses of the form $x_i$ can be used, for example in the
bilinear model $p_{ij}=x_i x_j $ that is associated to the sparse hypothesis
\cite{park_statistical_2004, bargigli_statistical_2014}.

The following functional form was introduced in \cite{garlaschelli_fitness-dependent_2004} 
to reconstruct the World Trade Web using the Gross
Domestic Product (GDP) of various countries as an explanatory variable: 
\begin{eqnarray}
 p_{ij} &=& \frac{z ~x_i x_j}{ 1+ z~x_i x_j }
 \label{eq.pij.ratio.x_i.x_j}
\end{eqnarray}
This form is an example of ``fitness-induced configuration model'' (FiCM), 
that makes it possible to respect both the maximum likelihood criterion
and the total empirical number of links, or equivalently the overall
link density of the network, as noted in \cite{garlaschelli_maximum_2008}.


Many extensions were debated to cover for example weighted, directed,
bipartite graphs. In the next section, dyadic terms such as the distance $d_{ij}$ 
in the gravity model $p_{ij}=\frac{x_i x_j}{ d_{ij}^2}$, 
where the respective contributions of nodes $i$ 
and $j$ can't be disentangled, will be introduced.

\subsection{Application to a disaggregated SFC model}
\label{sub.application.network.bmw}

In sec. \ref{sec.sfc.model} an example of disaggregated economic model was
considered. In sec. \ref{sec.network.reconstruct}, reconstruction
strategies were discussed. In the present section, the nature of the
graph built according to the model in sec. \ref{sec.sfc.model} is explained,
as well as the
probabilistic models used to reconstruct it from partial data.
Section \ref{sub.empirical.data} describes the necessary data.

Let $G$ be the graph associated to the transactions of the
disaggregated model. It is composed of various subgraphs with heterogeneous
properties, due to the diversity of the transactions involved, as
summarized in Tab. \ref{tab.subgraphs}, where all subgraphs are binary
and do not include any self-loop (a similar graph representing assets
and liabilities may be written, following \cite{bardoscia_pathways_2017}).

$G$ is a multipartite, multilayer  network. The population of agents is
divided into three categories (banks, firms, households). Each type of
transaction corresponds to one layer. This type of network, while common
for social networks \cite[Tab.2]{kivela_multilayer_2014}, is not often
found in economic empirical studies that tend to focus on individual
layers, one at a time (finance, interbank, production, consumption, \ldots). 
Furthermore, imposing a constraint over the network (here $A\xi=b$)
is used in engineering to reflect conservation laws (mass, energy),
or in statistical mechanics of networks, for example to study
motif-constrained ensembles (e.g. 2-stars, see
\cite{park_statistical_2004}), but is not commonly found in economic
network models.

\begin{table}[htbp]
\centering
	\begin{tabular}{lllp{3.5cm}} %
	\hline
	Transaction type & Graph type & Nodes & Edge $i\rightarrow j$ or $i\leftrightarrow j$ present if: \\	
	\hline
	Investment of firms & unipartite directed & firms &  firm $i$ is selling capital goods to firm $j$ \\
	Consumption of households& bipartite undirected  & firms, households &  firm $i$ is selling consumption goods to household $j$ \\
	Wages & bipartite undirected & firms, households & firm $i$ pays a wage to household $j$\\
	Interests on loans & bipartite undirected & banks, firms &  bank $i$ gets interests from firm $j$ \\
	Interests on deposits & bipartite undirected  & banks, households & bank $i$ pays interests to household $j$ \\
	\end{tabular}
\caption{\label{tab.subgraphs} Graph type for all subgraphs of the full transaction graph $G$. All subgraphs are binary.}
\end{table}

Following the methods explained in sec. \ref{sec.network.reconstruct}, 
we need to specify
the probability model $p_{ij}$ for each type of transaction using the different ingredients
available (node-specific or dyadic terms, \ldots).
Not all subgraphs could be modelled as FiCM maximum-entropy networks, due to the 
availability of data to fit the models. In that case, BiRG networks were used.

\begin{figure}[htbp]
\centering
\subfigure[~]{
	\includegraphics[width=4cm]{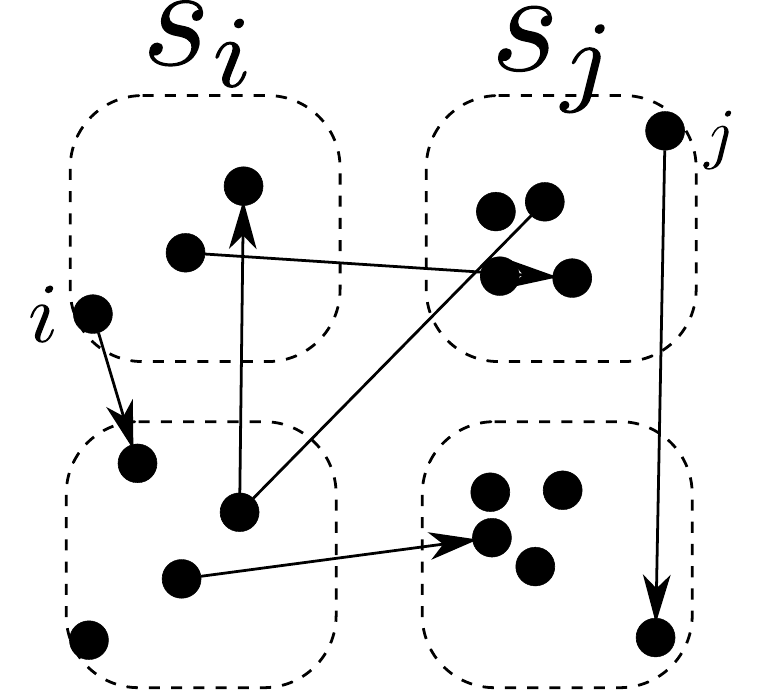}}
\subfigure[~]{
	\includegraphics[width=6cm]{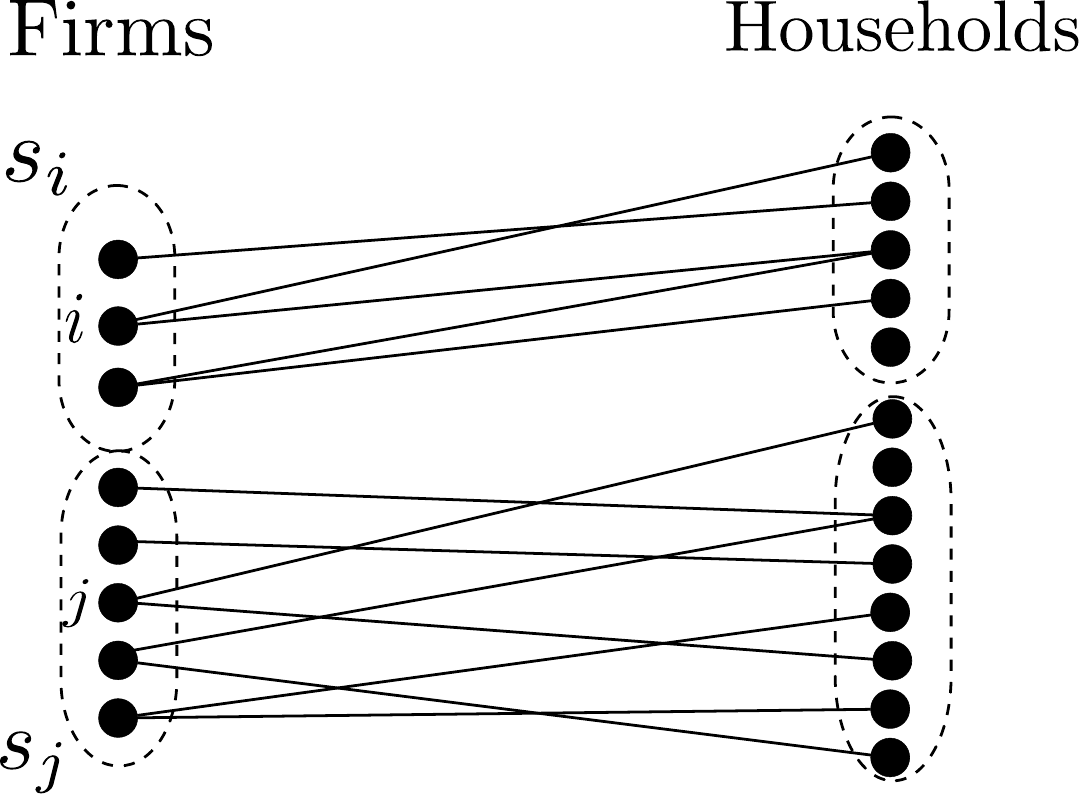}}
\caption{Simplified network of (a) investments between firms $i$ and $j$ belonging
to different sectors $s_i$ and $s_j$ ; (b) wages.}
\label{fig.ficm}
\end{figure}

Let us first remark that in a few cases, the available empirical data
impose a community structure to the graph. For example, the demography of
firms in public datasets is given at an aggregate sectorial level.
The same is true for intermediate consumption and the investment in capital
goods of firms (see sec. \ref{sub.empirical.data}).
As a consequence, the probability matrix $\{ p_{ij} \}$ that firm investment
occurs between any two firms $i$ and $j$ will in the simplest model
be a block matrix with terms taken in $\{ p_{s_i,s_j} \}$, the 
probability matrix that a transaction occurs between sectors $s_i$ and $s_j$, 
as represented in Fig. \ref{fig.ficm}(a). The same is true for consumption
of households, because the source of final consumption of households is
known at the level of industrial sectors.
 
The different functional forms corresponding to the transactions 
of this simplified block model are summarized 
in Tab. \ref{tab.pij.block}. $z$ is a free parameter which value is set independently
for each transaction using maximum likelihood, as will be seen below. 
Dyadic factors are noted $d_{s_i s_j}$ and node-specific factors are noted $x_i$. 
Using node-specific rather than dyadic factors in FiCM is also a matter 
of data availability, as will
be seen in sec. \ref{sub.empirical.data}.

\begin{table}[htbp]
\centering
	\begin{tabular}{p{4cm}lllp{4cm}} %
	\hline
	Transaction type & Model type & $p_{ij}$ & Indices & Constraint \\	
	\hline
	Investment of firms & FiCM &$\frac{z ~d_{s_i s_j}}{1+ z ~d_{s_i s_j}}$  & $(i,j) \in [1,nf]^2$ & $d_{s_i s_j}$: propensity of sector $s_i$ to sell capital goods to $s_j$    \\
	Consumption of households & FiCM & $\frac{z ~x_{s_i}}{1+z ~x_{s_i}}$ &  $(i,j) \in [1,nf]\times [1,nh]$  & $x_{s_i}$: propensity of sector $s_i$ to sell consumption goods to households   \\
	Wages & BiRG  & $\frac{L}{nf ~nh}$ &  $(i,j) \in [1,nf]\times [1,nh]$ & $\langle L \rangle=L^*$, or $\langle k \rangle=k^*$ \\
	Interests on loans & BiRG & $\frac{L}{nb ~nf}$ &  $(i,j) \in [1,nb]\times [1,nf]$ & $\langle L \rangle=L^*$, or $\langle k \rangle=k^*$  \\
	Interests on deposits & BiRG & $\frac{L}{nb ~nh}$ &  $(i,j) \in [1,nb]\times [1,nh]$ & $\langle L \rangle=L^*$, or $\langle k \rangle=k^*$  \\
	\end{tabular}
\caption{\label{tab.pij.block} Block model for each transaction type. All subgraphs 
are binary. $z$ is a free parameter which value is set independently for each
transaction using maximum likelihood. }
\end{table} 

\begin{table}[htbp]
\centering
	\begin{tabular}{p{4cm}lllp{4cm}} %
	\hline
	Transaction type & Model type & $p_{ij}$ & Indices & Constraint \\	
	\hline
	Investment of firms & FiCM & $\frac{z a_i a_j ~d_{s_i s_j}}{1+ z a_i a_j ~d_{s_i s_j}}$ & $(i,j) \in [1,nf]^2$ &  $a_i$: propensity of firm $i$ to play a role in investment, either as a buyer or a seller.    \\
	Wages & FiCM  & $\frac{z a_i x_{s_i}}{1+ z a_i x_{s_i}}$ & $(i,j) \in [1,nf]\times [1,nh]$  & $a_i$: see investment. $x_{s_i}$: propensity of sector $s_{i}$ to attract the workforce \\
	\end{tabular}
\caption{\label{tab.pij.rndfitness} Random fitness model for transactions that differ from the block model in Tab. \ref{tab.pij.block}. }
\end{table}

The first two rows in Tab. \ref{tab.pij.block} are associated with FiCM networks
which definitions were recalled in sec. \ref{sec.network.reconstruct}.
In the case of consumption, all households are considered homogeneous.
This explains the difference in the form of $p_{ij}$ between the first
two lines.

The mean number of links 
$\langle L \rangle = \sum_i \sum_{j \neq i} p_{ij}$ in the undirected case
is constrained and should be equal to the observed one 
$L = \sum_i \sum_{j \neq i} a_{ij}$, which is given by empirical observation. 
Under the chosen models, given empirical fitnesses and $L$, the values of the remaining 
free parameters $z$ for each network is set using maximum-likelihood procedure, for
the probabilities $p_{ij}$ to be fully specified.
In the case of the FiCM model for investments, this leads to solving the one-dimensional
nonlinear equation in $z$:
\begin{eqnarray}
 L &=& \sum_{i<j} \frac{z ~x_i x_j}{1+z ~x_i x_j}
\end{eqnarray}
which can be done approximately using standard numerical methods.
One can then sample independently the edges, and obtain random network samples for each
transaction. An example is given in Fig. \ref{fig.topological.properties}(d).

In Tab. \ref{tab.pij.block} the last three rows refer to a BiRG network (which
definition is recalled in sec. \ref{sec.network.reconstruct}) such that
only the average number of links is constrained. This type of model is not
appropriate if local information is available, but this is not the case
for the examined transactions. The edge probability is uniform and has the
simple expression $p_{ij}=\frac{L}{N_1 N_2}$ where $N_1$ and $N_2$ are
the number of vertices in each layer of the bipartite network.
The average number of links $\langle L \rangle=L^*$ constraint can
be equivalently expressed as a mean degree constraint, either on one subset of the nodes
or the other since these networks are bipartite with a fixed number of vertices.
Furthermore, for the BiRG networks: 
\begin{itemize}		
\item Wages: this network (firms on one hand and households on the other)
can be subdivided into different independent bipartite networks that
correspond to industrial sectors as shown in Fig. \ref{fig.ficm}(b). Firms 
and households are randomly assigned to a sector depending on business demography data 
(i.e. number of firms and number of employees in each sector). Then the connection
probabilty is set such that the mean degree on the households side equals
the empirical mean: 
	$\forall i \in [1,nh], ~\langle k_i \rangle=\bar{k}^{wage}$
\item Interests on loans: the degree of this network (banks and firms)
is uniform on the firms side, and equals its empirical mean: 
	$\forall i \in [1,nf], ~\langle k_i \rangle=\bar{k}^{loans}$
\item Interests on deposits: the degree of this network (banks and 
households) is uniform on the households side, and equals its empirical mean: 
$\forall i \in [1,nh], ~\langle k_i \rangle=\bar{k}^{deposit}$
\end{itemize}			

In the block model, for some subnetworks, all firms in a given economic sector
have the same connection probability with firms in another sector. This 
approach is useful but not realistic. A simple way to introduce diversity among
sectors is to use random fitnesses sampled from a specific distribution. 
Instead of writing $p_{ij}$ in the investment network in the form
$\frac{z ~d_{s_i s_j}}{1+ z ~d_{s_i s_j}}$, one may write:
\begin{eqnarray}
p_{ij} &=& \frac{z a_i a_j ~d_{s_i s_j}}{1+ z a_i a_j ~d_{s_i s_j}}
\label{eq.pij.invest.rndfitness}
\end{eqnarray}
where $a_i$ are firm-specific terms sampled from some distribution
that can be fit to empirical values. In this article a continuous uniform
distribution $X=\mathcal{U}_{[0,1]}$ is chosen, but power law distributions
will be considered in further works. The wage network is modified accordingly,
and can be seen as an FiCM network determined by the firm-specific fitness 
$a_i$ in eq.(\ref{eq.pij.invest.rndfitness}) and a sector-specific fitness
$x_{s_i}$ that reflects the sectorial
assignment of the workforce in empirical data, keeping the household sector
homogeneous:
\begin{eqnarray}
p_{ij} &=& \frac{z a_i x_{s_i}}{1+ z a_i x_{s_i}}
\end{eqnarray}
The specification of this random fitness model is 
summarized in Tab.\ref{tab.pij.rndfitness}. Other propositions to go beyond
the limits of this formalization will be discussed in sec.\ref{sec.discussion}

\subsection{Empirical data and fitnesses}
\label{sub.empirical.data}

Most datasets are extracted from Eurostat databases 
\cite{eurostat_:_statistisches_amt_der_europaischen_gemeinschaften_eurostat_2008} 
as summarized in Tab. \ref{tab.datasets}. The figures mentionned in this table
concern the demography of businesses and are not used directly to initialize 
simulations, but are downscaled to permit computation in reasonable time.
The agricultural sector, being separated from business demography data, is not 
included in this study but will be added in further works.

The structure of supply, use and input-output tables is explained in \ref{appendix.supply.use},
along with the notations for the Eurostat sector aggregates.
Each correspond to a given country, and a given year, but these 
mentions are dropped for simplicity.

These datasets are used to compute the fitnesses $d_{i,j}$ and $x_{i}$ in
Tab. \ref{tab.pij.block}. $d_{s_i s_j}$ quantifies the propensity of
sector $s_i$ to sell capital goods to $s_j$.
Since this information is not available directly in public datasets, our
proposition is to use the volume of intermediate consumption by industry 
in the use table as a proxy for $d_{s_i s_j}$.
It is a dyadic factor depending specifically on the couple $(s_i,s_j)$.  

The fitness value $x_{s_i}$ in Tab. \ref{tab.pij.block} is supposed to quantify
the propensity of sector $s_i$ to sell consumption goods to households.
While household consumption could in fact be separated into different
subgroups (see \cite{leo_correlations_2018}), we keep it aggregated here
for simplicity. The proposed proxy comes in the form of a dot-product:
\begin{eqnarray}
 x_{s_i} & \propto &  \sum_{p \in [1,P]} sup[p,s_i] \times use^{fin}[p]
\end{eqnarray}
where $sup[p,s]$ is the value of product $p \in [1,P]$ produced by sector
$s \in S$, and $use^{fin}[p]$ is the value of product $p \in [1,P]$ consumed by 
households as a final use. This factor weighs the proximity between the
supply profile of sector $s_i$ and the final use profile of the household
sector. It is normalized to $1$ across sectors.

Average degrees of networks typically are not included in public datasets, 
but rather in private ones. However their magnitude can be looked for in
the literature in order to parametrize both FiCM and BiRG models in sec. \ref{sub.application.network.bmw}:

\begin{itemize}		
\item the average number of edges per vertex for the interfirm network
has been studied in the USA for the period 1979-2002, and
has a average value of $1.06$ according to \cite{atalay_network_2011}. 
In the case of Japan, the estimated value is close to 5 \cite{mizuno_buyer-supplier_2015}.
We believe these estimations aggregate both intermediate consumption of
firms, and acquisition of capital goods by firms. 
However we consider they are valuable proxies for the mean link number
$\langle L \rangle$ in the investment network.
\item the average number of suppliers for households can be estimated
from recent studies at the individual level \cite{dong_social_2017}.
We set it to 20 in experiments below.
\item the average number of jobs per household $\bar{k}^{wage}$ is set to $1$.
\end{itemize}
These choices are further discussed in sec.\ref{sec.discussion}.

\subsection{Network properties} 
\label{sub.valid.topo}

In this section we analyze the properties of some of the network
models discussed in sec. \ref{sub.application.network.bmw}. The 
availability of a probabilistic model
allows us to compute analytically the moments of some higher-order
topological properties \cite{squartini_maximum-entropy_2017},
so as to compare it to stylized facts found in empirical studies.
The programs used to generate the figures are publicly available\footnote{\url{https://gitlab.com/hazaa/sfc_proba}}.
The figures dealt with in this section concern the Czech Republic in 2010.

\begin{figure}[htbp]
\centering
\subfigure[~]{
	\includegraphics[width=8cm]{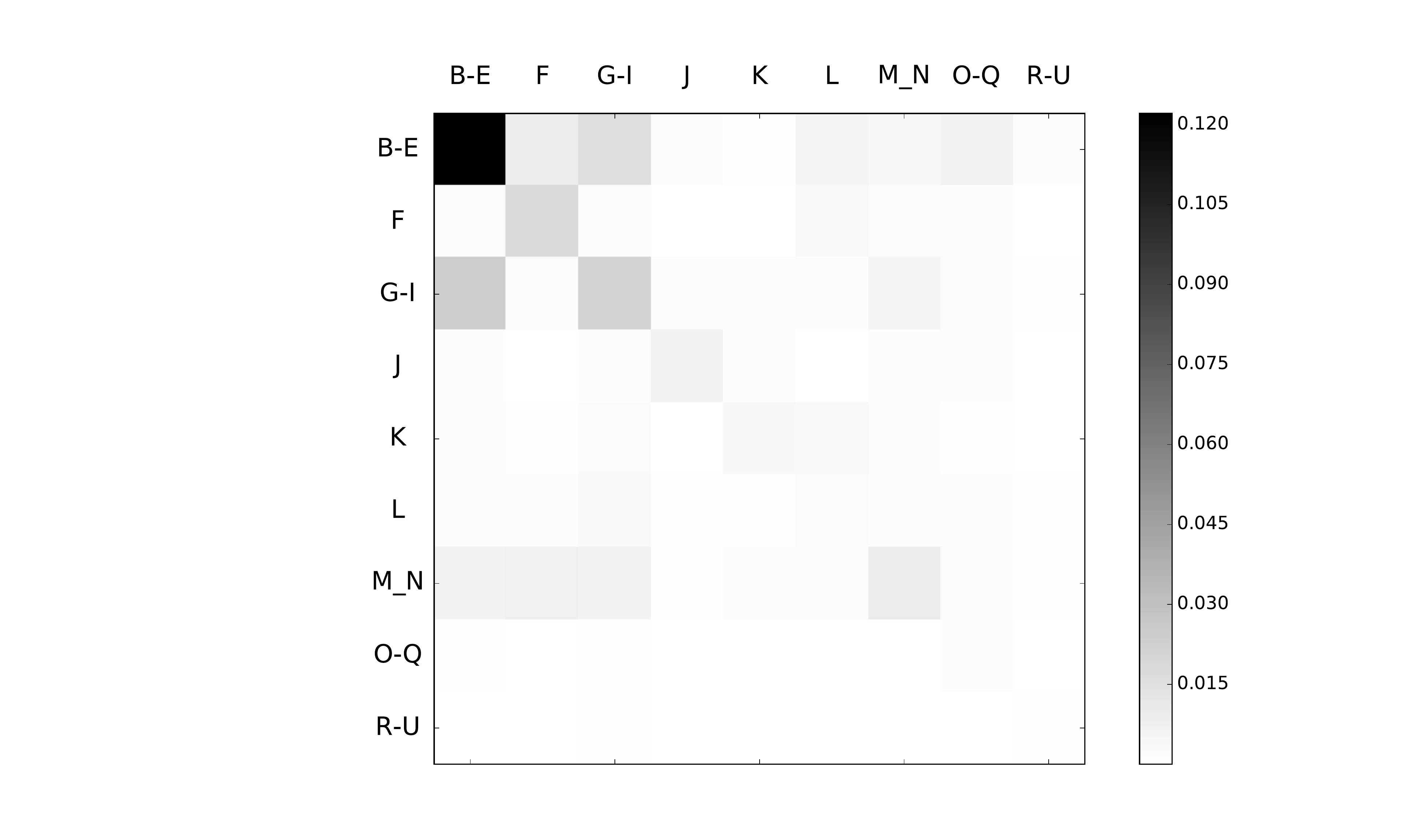}}
\subfigure[~]{
	\includegraphics[width=8cm]{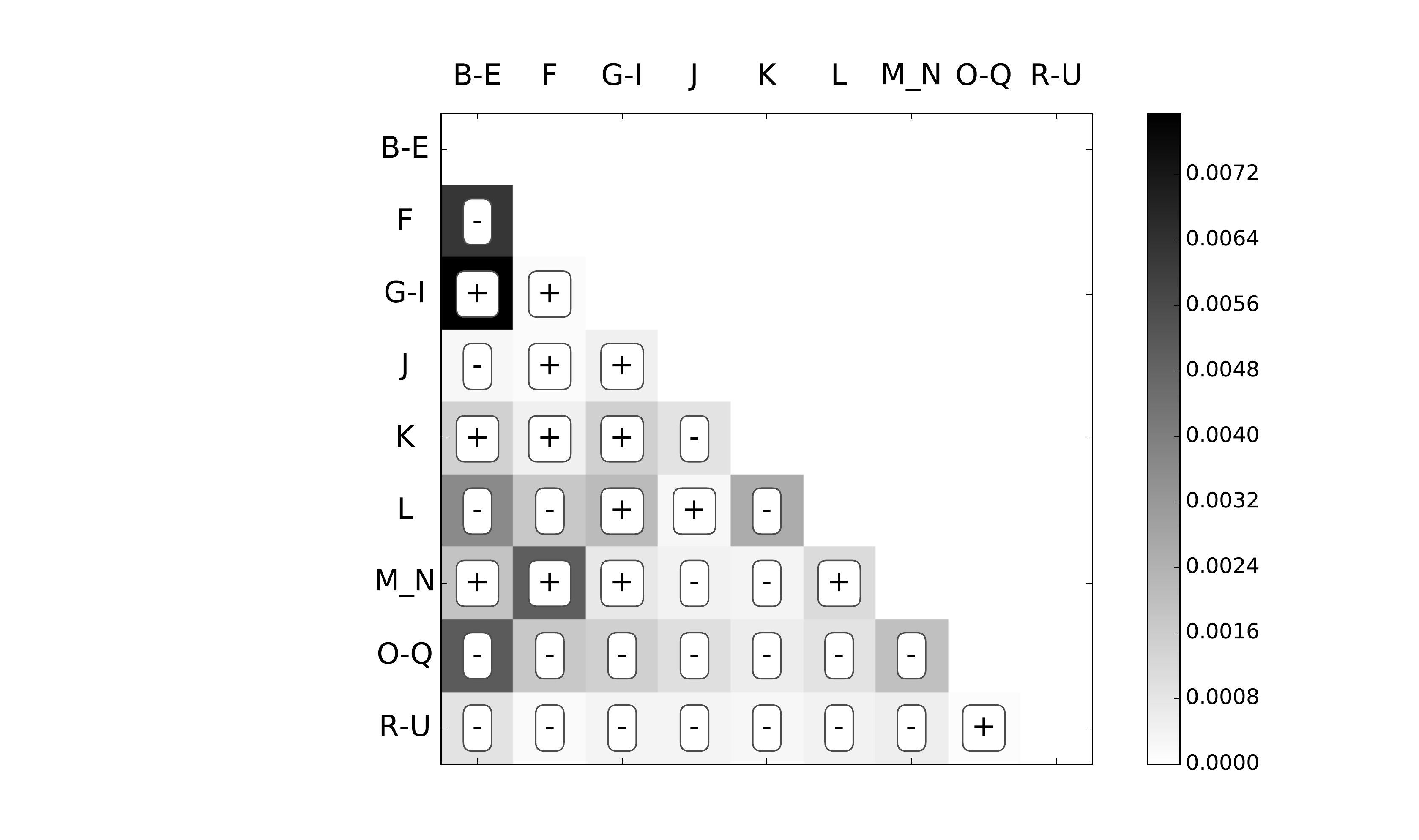}}	
\subfigure[~]{
	\includegraphics[width=8cm]{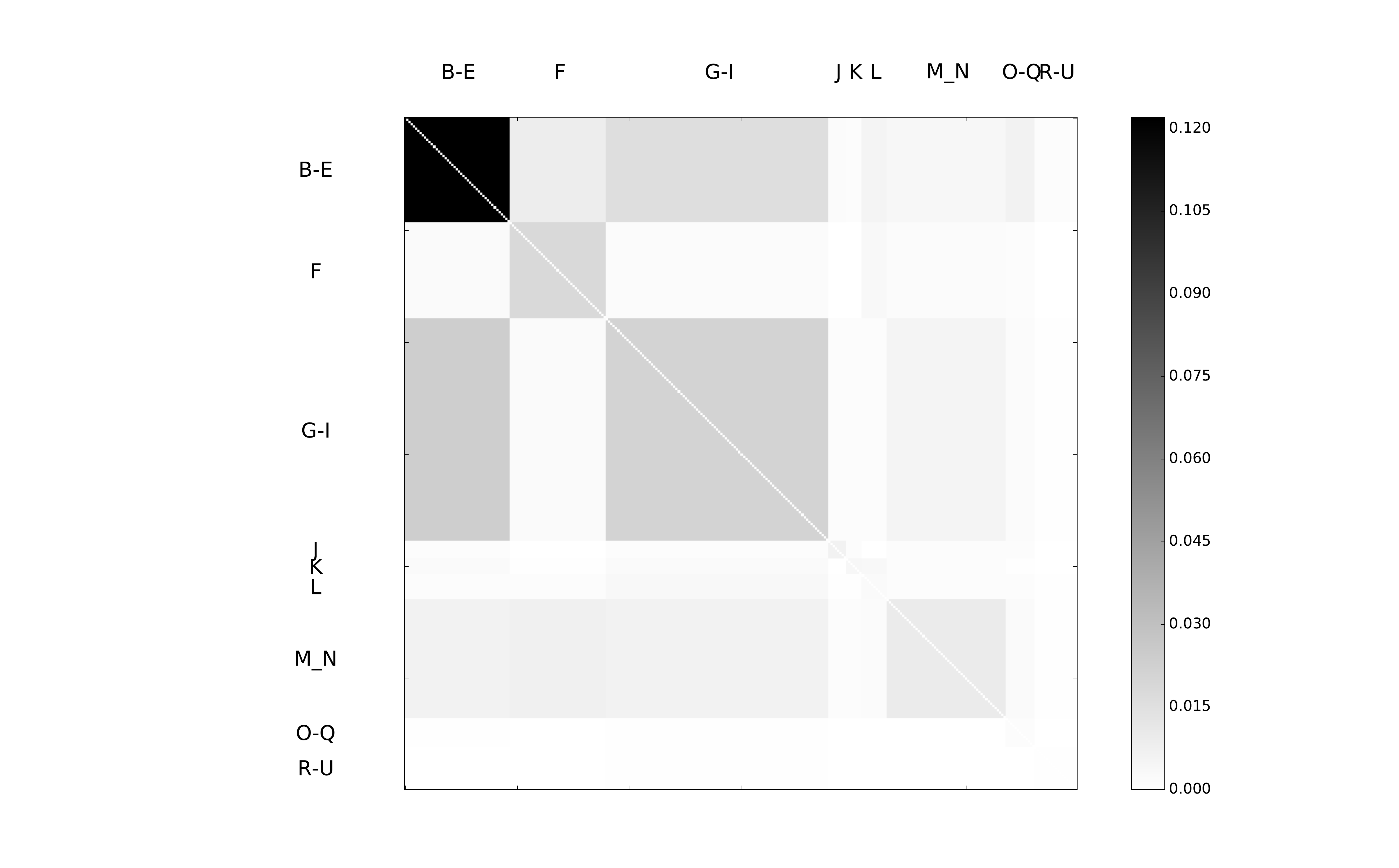}}
\subfigure[~]{
	\includegraphics[width=7cm]{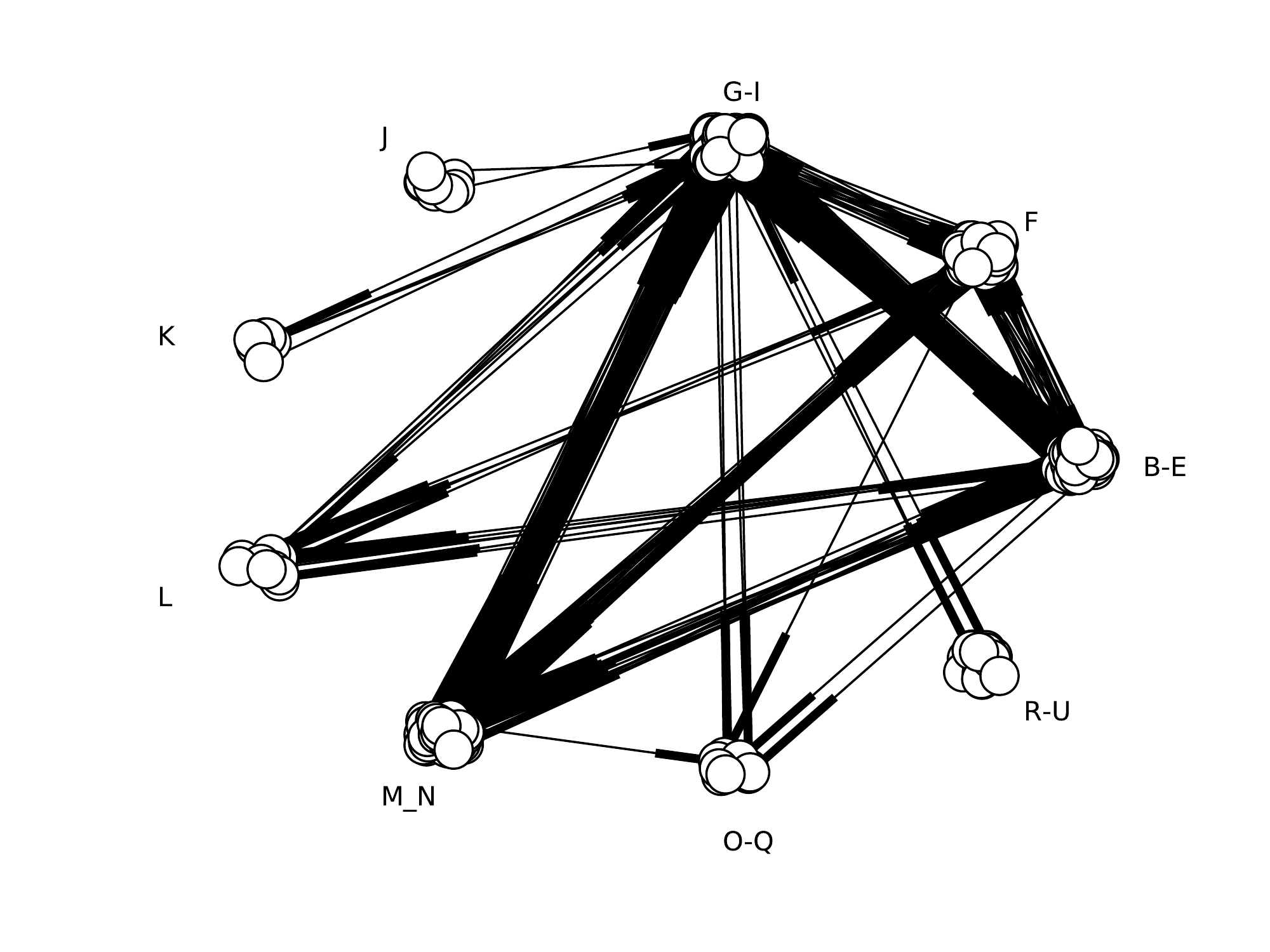}}	
\caption{Properties of the investment block network, unipartite
and directed. (a) sector-level connection probability matrix $p_{s_i,s_j}$  
(b) lower triangle part of the matrix $\{ |p_{s_i,s_j}-p_{s_j,s_i}| \}$
with the corresponding sign in the bounding box 
(c) probability matrix $p_{ij}$ with $nf=300$ 
(d) random network sample.}
\label{fig.topological.properties}
\end{figure}

In sec. \ref{sub.application.network.bmw}, the expression of the connection
probabilities was given for all transaction networks.  
Fig. \ref{fig.topological.properties}(a) illustrates the behavior of 
$p_{s_i,s_j}$ in the case of the investment block network, proxied by the 
interfirm consumption network in this article. 
Almost all sectors are primarily connected to themselves. Four 
aggregated sectors stand out in terms of probability magnitude: B-E, 
F, G-I, and M-N (see definitions in Tab. \ref{tab.sector.labels}). This can
be related to the number of firms in each sector, jointly represented
with probabilities in Fig. \ref{fig.topological.properties}(c).
In first approximation the latter can be considered as symmetric, 
but this is not strictly verified at a finer scale as shown in 
Fig. \ref{fig.topological.properties}(b), where 
$\{ |p_{s_i,s_j}-p_{s_j,s_i}| \}$ is represented, as well
as the sign of the difference. For example industry (B-E) sells more to
construction (F) than the opposite.
A randomly generated network corresponding to $p_{ij}$ is shown in
Fig. \ref{fig.topological.properties}(d) and illustrates these facts. 

\begin{figure}[htbp]
\centering
\subfigure[~]{
	\includegraphics[width=8cm]{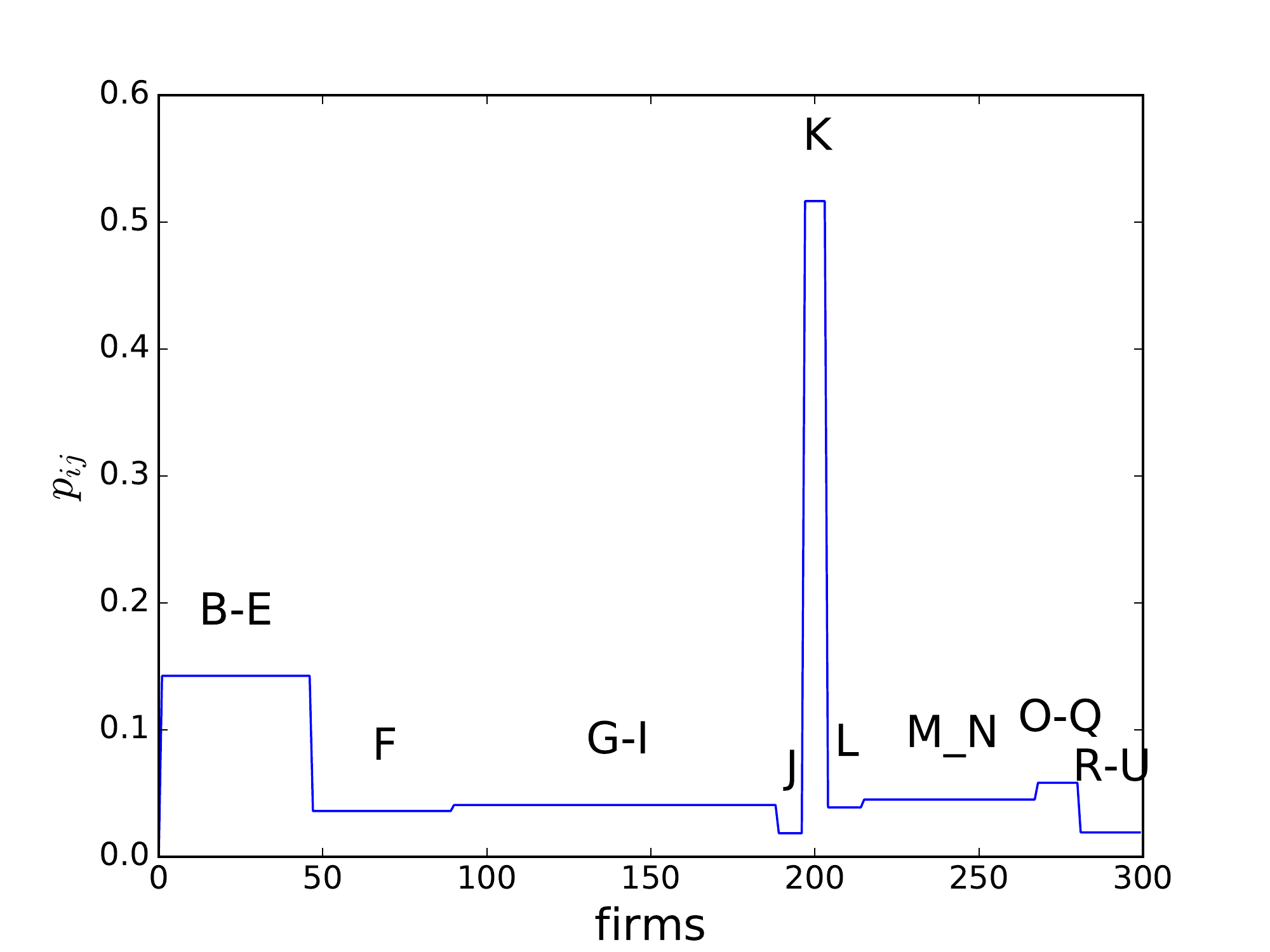}}
	\caption{Properties of the households' consumption block network, bipartite
and undirected. }
\label{fig.topological.properties.cons.hh}
\end{figure}

The network of households' consumption is simpler, as defined in 
sec. \ref{sub.application.network.bmw},
because the connection probability $p_{ij}$ doesn't depend on $j$.
Since the household sector is homogeneous, the matrix $p_{ij}$ can 
be summed-up by its cross-section along the firms' axis,
as shown in Fig. \ref{fig.topological.properties.cons.hh}.
Note the importance of sector K (``financial and insurance activities''). 

\begin{figure}[htbp]
\centering
\subfigure[~]{
	\includegraphics[width=8cm]{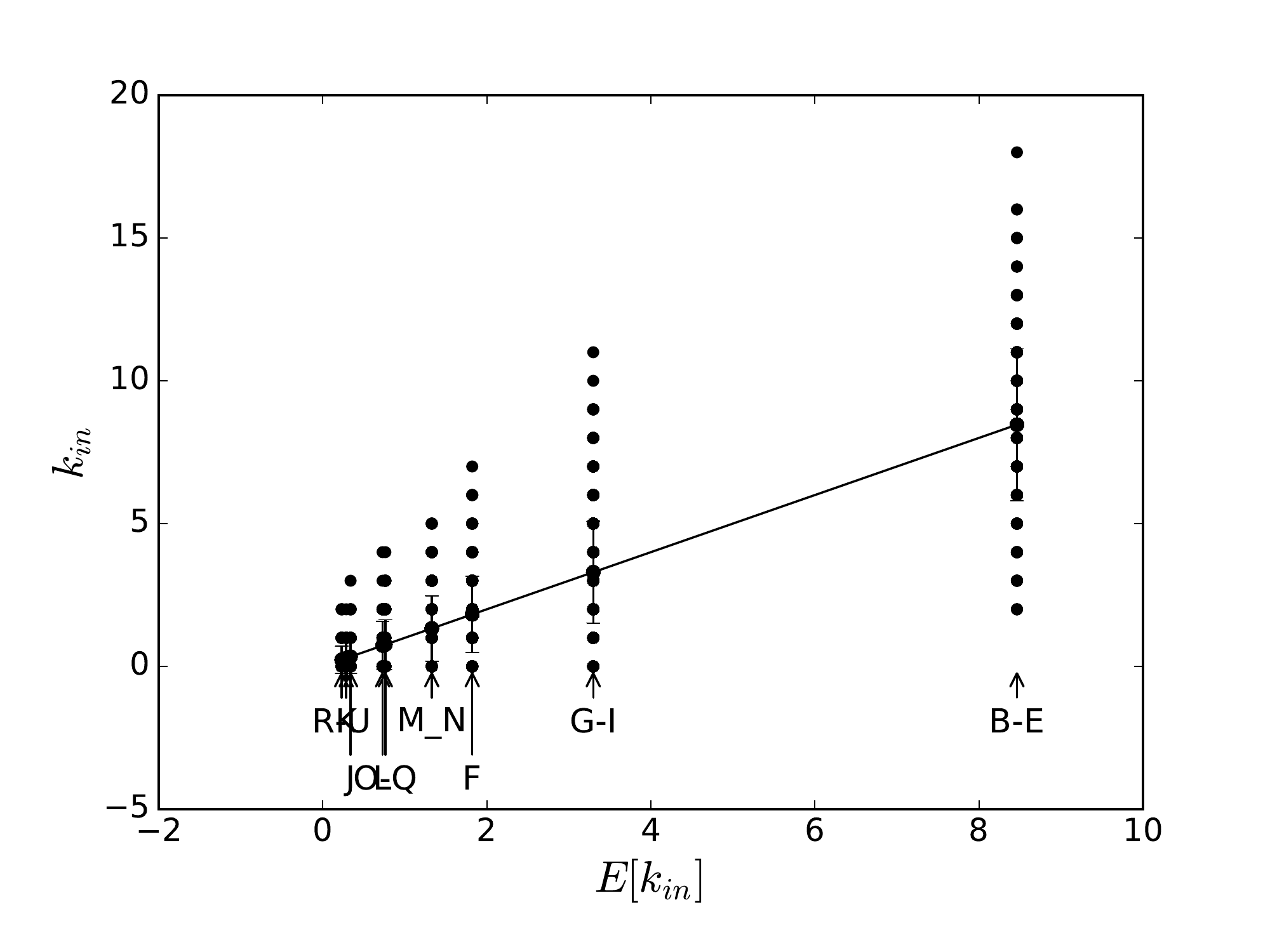}}
\subfigure[~]{
	\includegraphics[width=8cm]{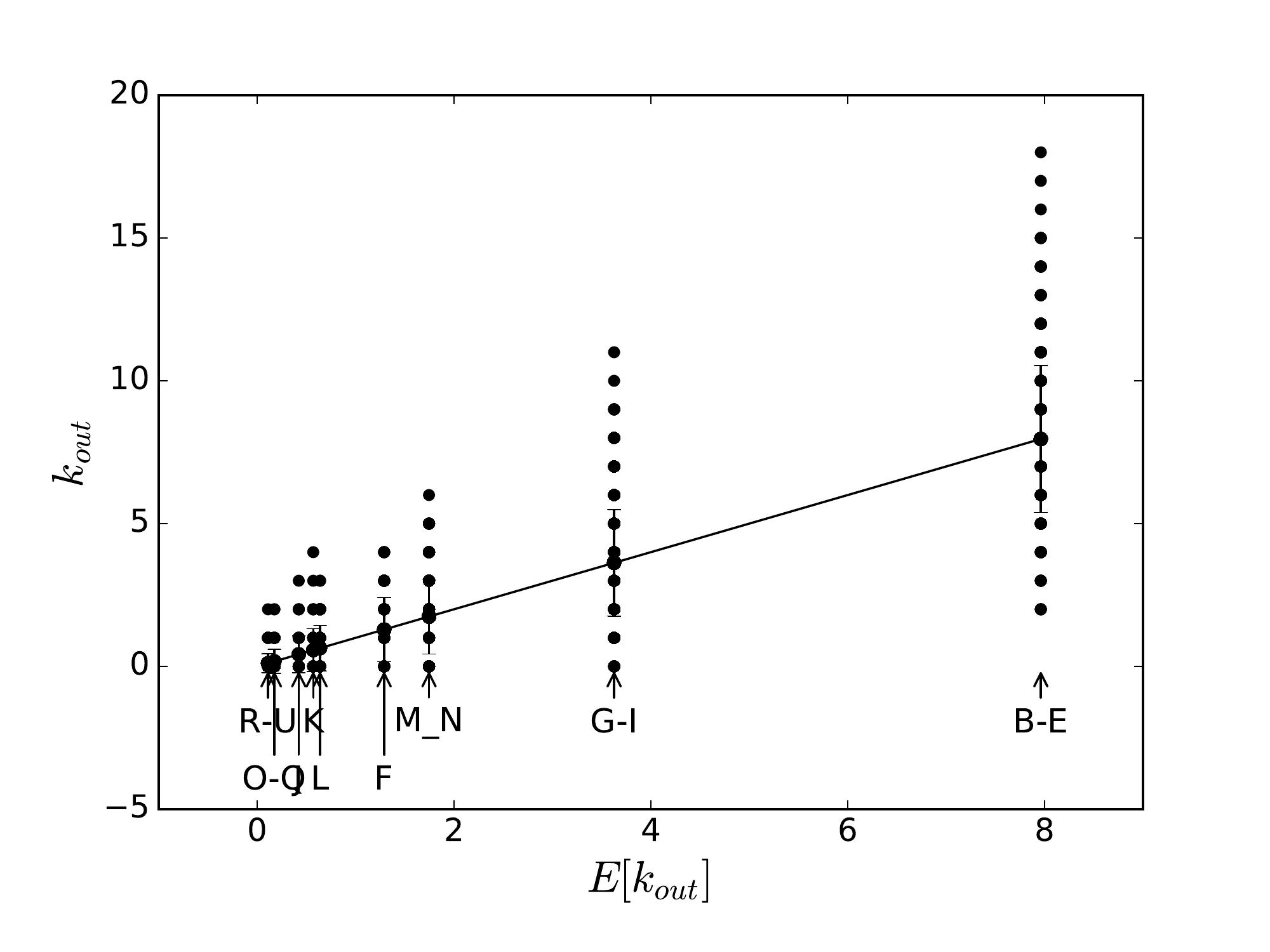}}
\caption{Degree of the investment block network. Dots are randomly sampled graphs,
lines represent theoretical approximation of the expectation. 
Errobar represent $\pm 1$ standard deviation (a) in-degree (b) out-degree. $nf=300$}
\label{fig.degree}
\end{figure}

The degree is another classical indicator of network topology.
Theoretical expectations of $k_{in}(j)$ and $k_{out}(i)$ are directly available
from $p_{ij}$, computing $\sum_i p_{ij}$ and $\sum_j p_{ij}$.
The variance can be computed using the fact that independent
Bernoulli variables are sampled along rows and columns. By a central
limit argument\footnote{more specifically the de Moivre-Laplace theorem.},
the pdf can be approximated for high degrees by a normal law.
Fig. \ref{fig.degree}(a-b) show sample degrees as functions of 
theoretical degree with approximate standard deviations plotted as error bars.
It can be noticed that, consistently with the block hypothesis,
all firms in a given sector have the same degree. Furthermore the four
prominent sectors are the same as in Fig. \ref{fig.topological.properties}.

Then we consider another classical high-order topological indicator,
the average nearest neighbors degrees. For each agent, it measures
the average degree of its neighbors.

\begin{eqnarray}
\label{eq.ann.degree}
k_{in}^{nn}(i) &=& \frac{ \sum_{j=1} a_{ij}k_{in}(j)}{k_{out}(i) } \\
k_{out}^{nn}(j) &=& \frac{ \sum_{i=1} a_{ij}k_{out}(i)}{k_{in}(j) }
\end{eqnarray}

\begin{figure}[htbp]
\centering
\subfigure[~]{
	\includegraphics[width=6cm]{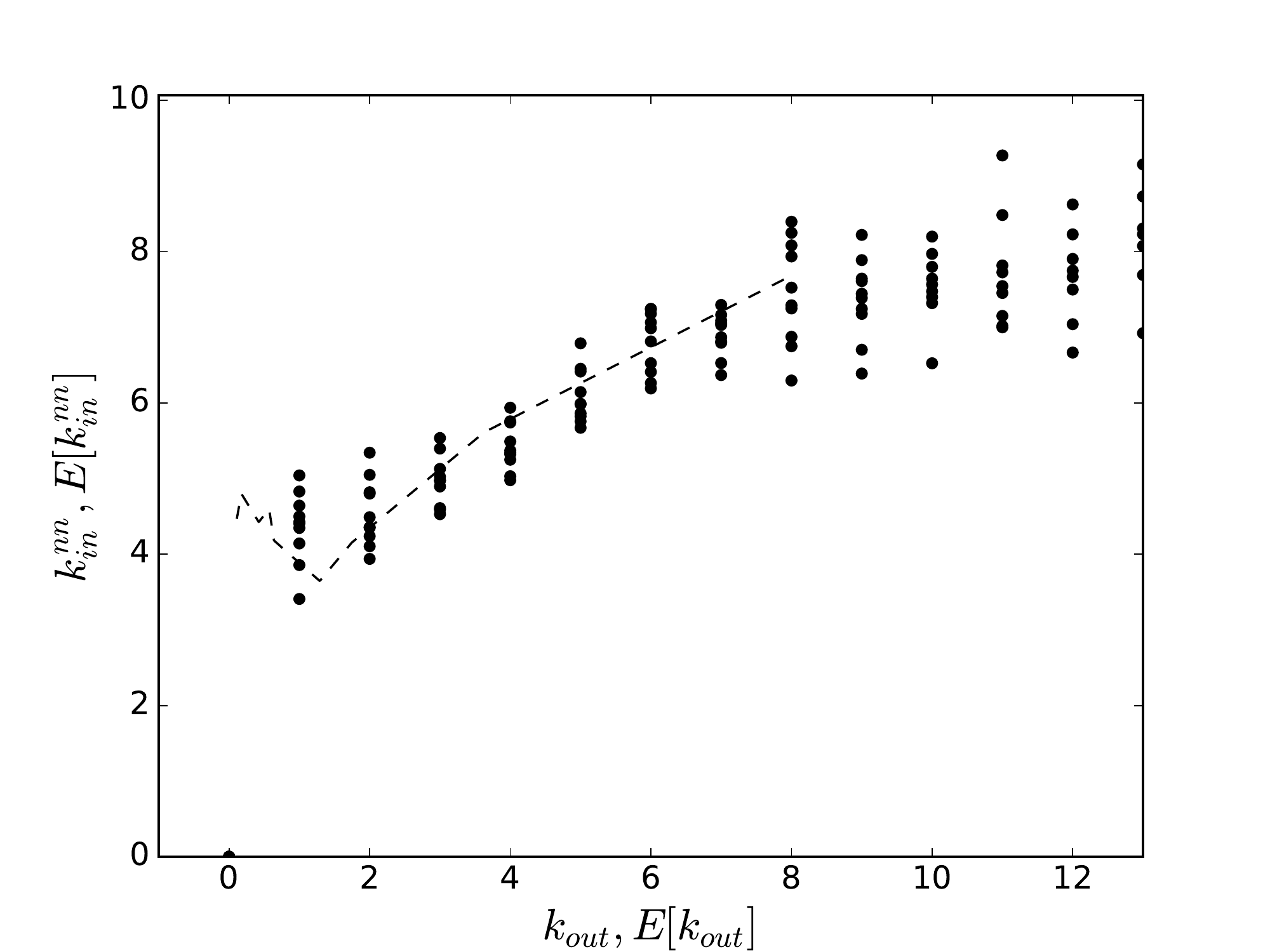}}
\subfigure[~]{
	\includegraphics[width=6cm]{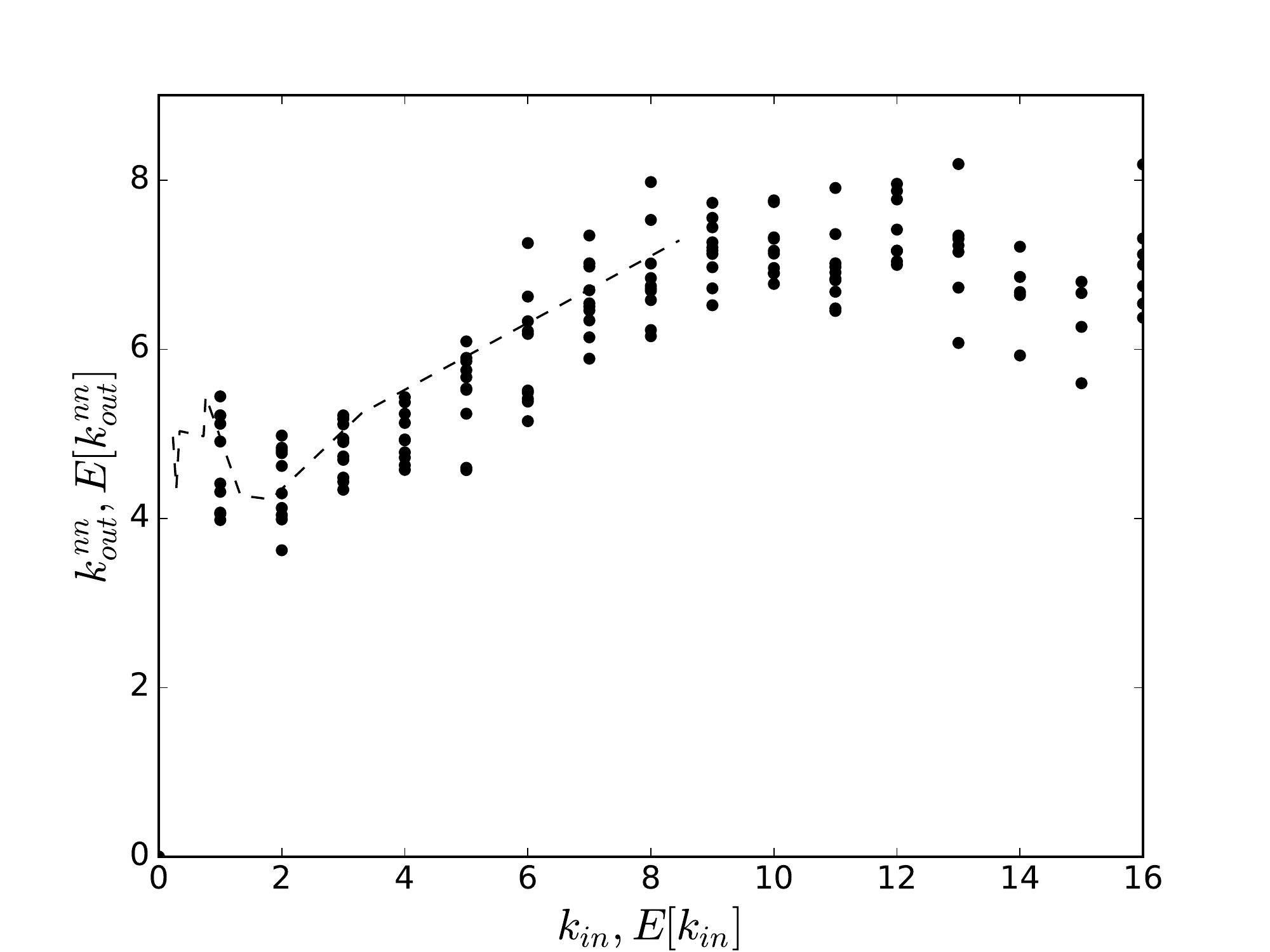}}
\caption{Nearest network degree of the investment block network. 
Dots are randomly sampled graphs, lines represent theoretical
approximations.}
\label{fig.ann.degree}
\end{figure}

Knowing $p_{ij}$, the expectation of these indicators can be approximated 
\cite[eq.(34-35)]{saracco_randomizing_2015}.
Using these results, in Fig. \ref{fig.ann.degree} we compare randomly 
sampled values and their theoretical couterpart.
Both agree concerning the assortative nature or the network, on the
buyers and suppliers sides: suppliers with a small (resp. high) out-degree
$k_{out}$ in Fig. \ref{fig.ann.degree}(a) tend to be connected to suppliers 
buyers with a small $k_{in}^{nn}$ (resp. high). The same is true for buyers 
with respect to in-degree. 
This is a consequence of the dense conection inside the first four
sectors (that account for 85\% of all links in this model),
be it with inter or intra-sector links. 
It contrasts with the disassortative nature of buyers-suppliers networks
found in recent empirical studies such as the interfirm payment network
in Estonia \cite{rendon_de_la_torre_topologic_2016}, 
Italy \cite[§2]{letizia_corporate_2018}, 
and Japan \cite[3.1]{bernard_production_2015}.

\begin{figure}[htbp]
\centering
\subfigure[~]{
	\includegraphics[width=8cm]{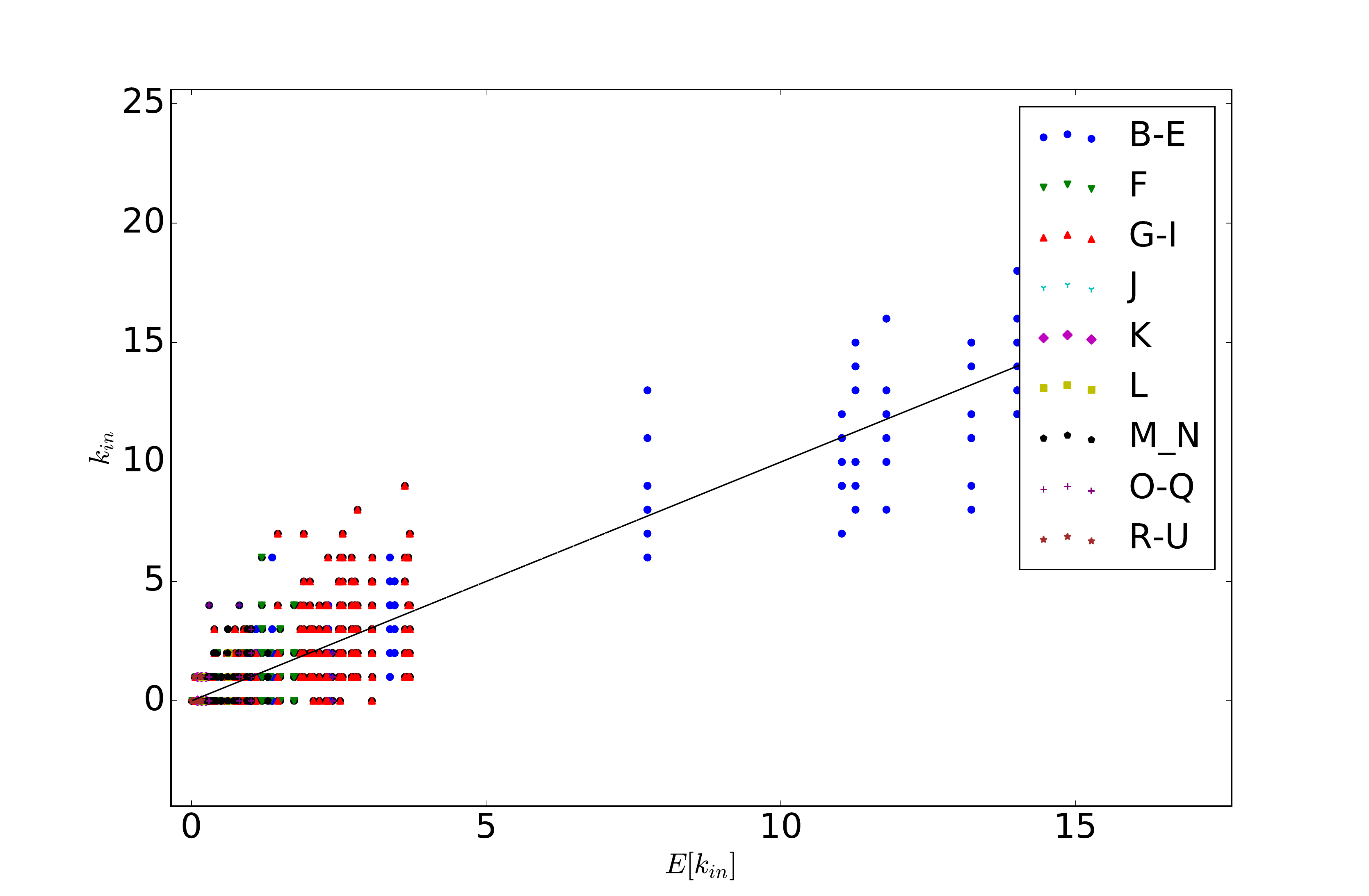}}
\subfigure[~]{
	\includegraphics[width=8cm]{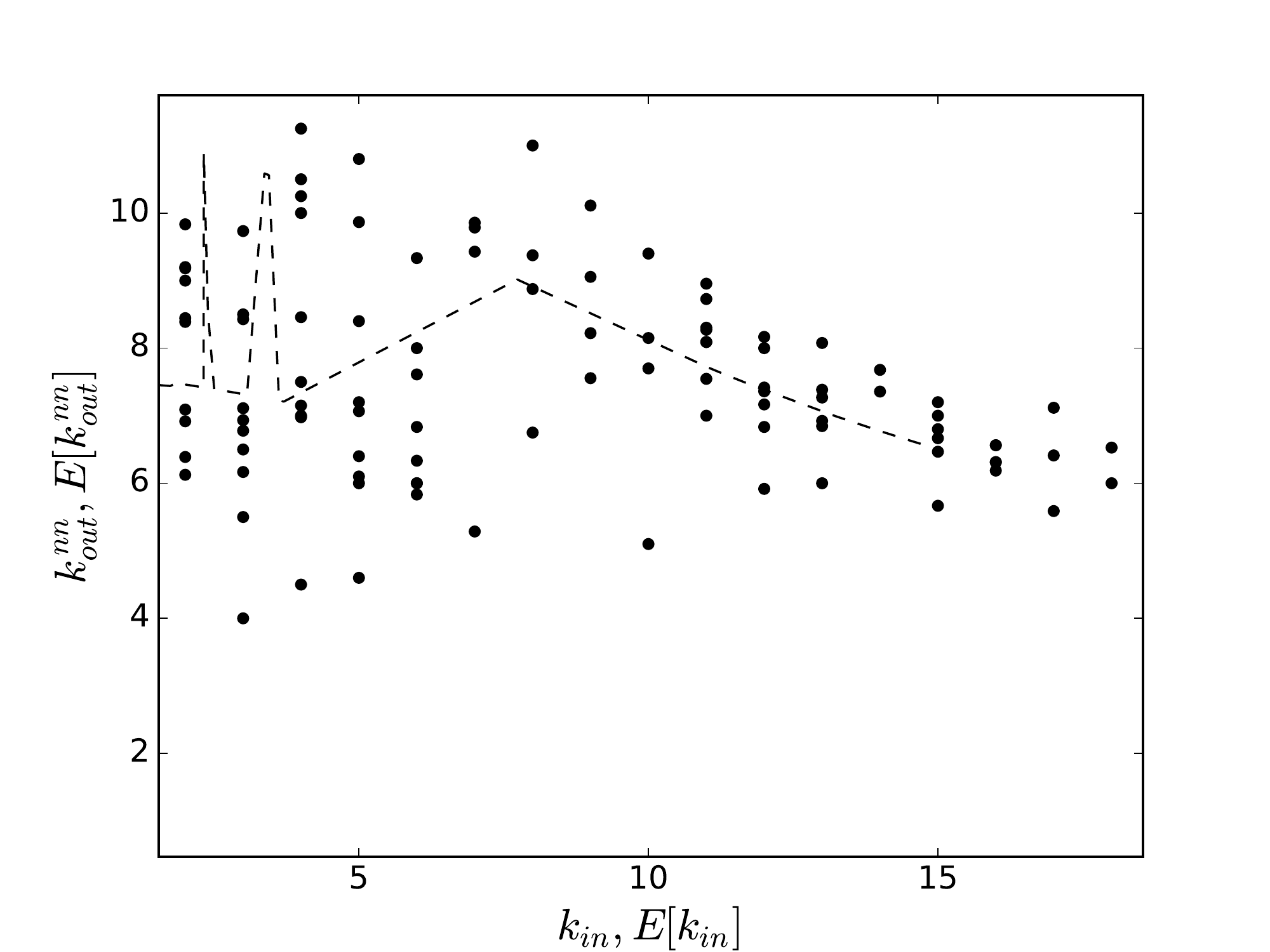}}
\caption{Random fitness investment network. (a) in-degree; (b) nearest-neighbour degree. $nf=300$}
\label{fig.degree.rndfitness}
\end{figure}

The random fitness model depicted in sec. \ref{sub.application.network.bmw}
and summarized in Tab. \ref{tab.pij.rndfitness} was proposed to 
avoid such inconsistencies. In Fig. \ref{fig.degree.rndfitness}(a) the 
in-degree distribution of the investment network shows that some
mixing was succesfully introduced among firms belonging to different
sectors, unlike in the case of block model in Fig.\ref{fig.degree} where
all firms of a given sector share the same degree.
Similarly, the investment network is no longer assortative in the
case of the random fitness network as shown by 
Fig. \ref{fig.degree.rndfitness}(b). 
It can be noticed that in that case, the probability distribution of 
degrees of a fitness model can be computed, following 
\cite[eq. (1-2)]{caldarelli_scale-free_2002}.

While in this section only topological features were examined, 
the topic of average money flows will also be dealt with below.



\section{Estimation of marginal probabilities of monetary flows for a particular network}
\label{sec.marginal.pdf}

In sec. \ref{sec.network.reconstruct} only the topological properties
of the model inspired by sec. \ref{sec.sfc.model} were examined.
The present section deals with the values of flows $\xi$ associated
to vertices given a specific topology. Moreover, we are 
interested in the relationship between the solutions and the topological
properties of networks.
The $n$-uple of adjacency matrices 
$\mathcal{A}_{SFC}=(\mathbf{A}_{cons},\ldots,\mathbf{A}_{invest})$ can be 
temporarily considered as a quenched variable: it is sampled once from the
statistical ensemble $\{ \mathcal{A}_{SFC} \}$ subject to empirical
constraints, and then considered fixed.

The linear time-invariant underdetermined problem in eq. (\ref{eq.syslin})
has the following properties: $A$ has full rank and is such that $m<n$ with $m$ the number of
rows and $n$ the number of columns. In that case, $AA^T$ is invertible,
the Moore-Penrose pseudoinverse writes $A^{\dag}=A^T(AA^T)^{-1}$, and
coincides with the least-square solution of eq. (\ref{eq.syslin}):
\begin{eqnarray}
\xi &=& A^T(AA^T)^{-1} b
\label{eq.pseudoinverse}
\end{eqnarray}

This result can be related to probabilistic approaches: Bayesian 
and maximum entropy 
methods were applied to flow networks in \cite{waldrip_comparison_2017},
using the unbounded Gaussian prior $\mathcal{N}(\mu,\Sigma)$. 
The authors get the following expression for the posterior average
\footnote{~to do so the likelihood is used to enforce the constraint
in eq.(\ref{eq.syslin}) $p(b|\xi) = \delta(b-A\xi)$, then
using the Gaussian expression of the delta function yields
$-2 \ln p(b|\xi) \propto \lim_{\Sigma_A \rightarrow 0}(b-A\xi)^T 
\Sigma_A^{-1}(b-A\xi) $} : 
\begin{eqnarray}
\langle \xi \rangle= \mu + \Sigma A^T ( A \Sigma A^T)^{-1} (b-A\mu)
\label{eq.solution.bayes}
\end{eqnarray}
With the simplifying homoscedasticity assumptions 
$\Sigma=\sigma I_n$, this solution can be compared to the one obtained
with the Moore-Penrose pseudoinverse in eq.( \ref{eq.pseudoinverse}).
However, several problems arise: 
firstly, eq.(\ref{eq.syslin}) has a trivial solution $x=0$ because
the system is homogeneous. It is thus necessary to turn to 
the nonhomogeneous system eq.(\ref{eq.syslin.alpha0}) to obtain
nontrivial solutions.
Secondly, negative solutions \footnote{~
for numerical reasons, the solution is not computed directly from
the expression of the pseudoinverse, instead sparse least-norm solvers
should be used.} may be found for this system, which limits the
interest of this method. 

Adding inequalities to eq.(\ref{eq.syslin}), it can be considered as a
Constraint Satisfaction Problem (CSP):
\begin{equation}
S=\{ \xi ~s.t. ~A \xi=b, \xi_0 \leq \xi \leq \xi_1 \}
\label{eq.CSP}  
\end{equation}
Various properties of the set of solution vectors $\xi$ 
can be studied numerically:
as examplified in the field of metabolic
networks research, 
samples can be generated in a uniform \cite{wiback_monte_2004, braunstein_estimating_2008}
or non-uniform way \cite{capuani_quantitative_2015}. This type of method was
applied already to SFC macro models in \cite{hazan_volume_2017}. 
Distributional properties were observed in the partially aggregated
case in \cite{hazan_stock-flow_2017}. 
Unfortunately, sampling methods can't be applied directly 
in the present case because of the dimension of the problem.
The positivity constraint can also be dealt with 
by the Expectation Propagation algorithm \cite{braunstein_analytic_2017}
which yields an analytic approximation of the marginal probability 
distribution $P(\xi_i)$, using truncated Gaussian priors.
But according to preliminary experiments the dimension of our problem 
is too large for existing implementations.

Other algorithms are then necessary to handle the constrained 
high-dimensional case.
As for metabolic network analysis \cite[§12.5]{palsson_systems_2006},
linear programming or basis pursuit can be used.
It the objective function to be minimized has the form $1^T\xi$,
sparse solutions can be found numerically. This property 
is interesting for example to force high unemployment, but is
too restrictive otherwise. 
Nonnegative Least Square (NNLS) numerically solves the problem:
\begin{equation}
  \argmin_{\xi} \| ~A \xi-b \|, s.t. ~\xi \geq 0 
\label{eq.nnls}  
\end{equation}
which is equivalent to a quadratic programming problem,
and which solution can be efficiently approximated 
\cite[§23]{lawson_solving_1995} for large problems.

\subsection{Properties of flows}
\label{sec.budget}

In this section we present numerical NNLS solutions to the
nonhomogeneous problem:
\begin{equation}
  \argmin_{\xi} \| ~A_1 \xi-b_1 \|, s.t. ~\xi \geq 0 
\label{eq.nnls.alpha0}  
\end{equation}
with $A_1,b_1$ defined in eq.(\ref{eq.syslin.alpha0}).
The networks that specify $A_1$ are sampled from the ensemble
defined by the random fitness model in sec. \ref{sub.application.network.bmw}.
Due to computational constraints, the size of the networks will be
limited to $n_b=3, ~n_f=100, ~n_h=1000$, resulting in a system with
more than $2.10^5$ unknowns.

\begin{figure}[htbp]
\centering
\subfigure[~]{
\includegraphics[width=7cm]{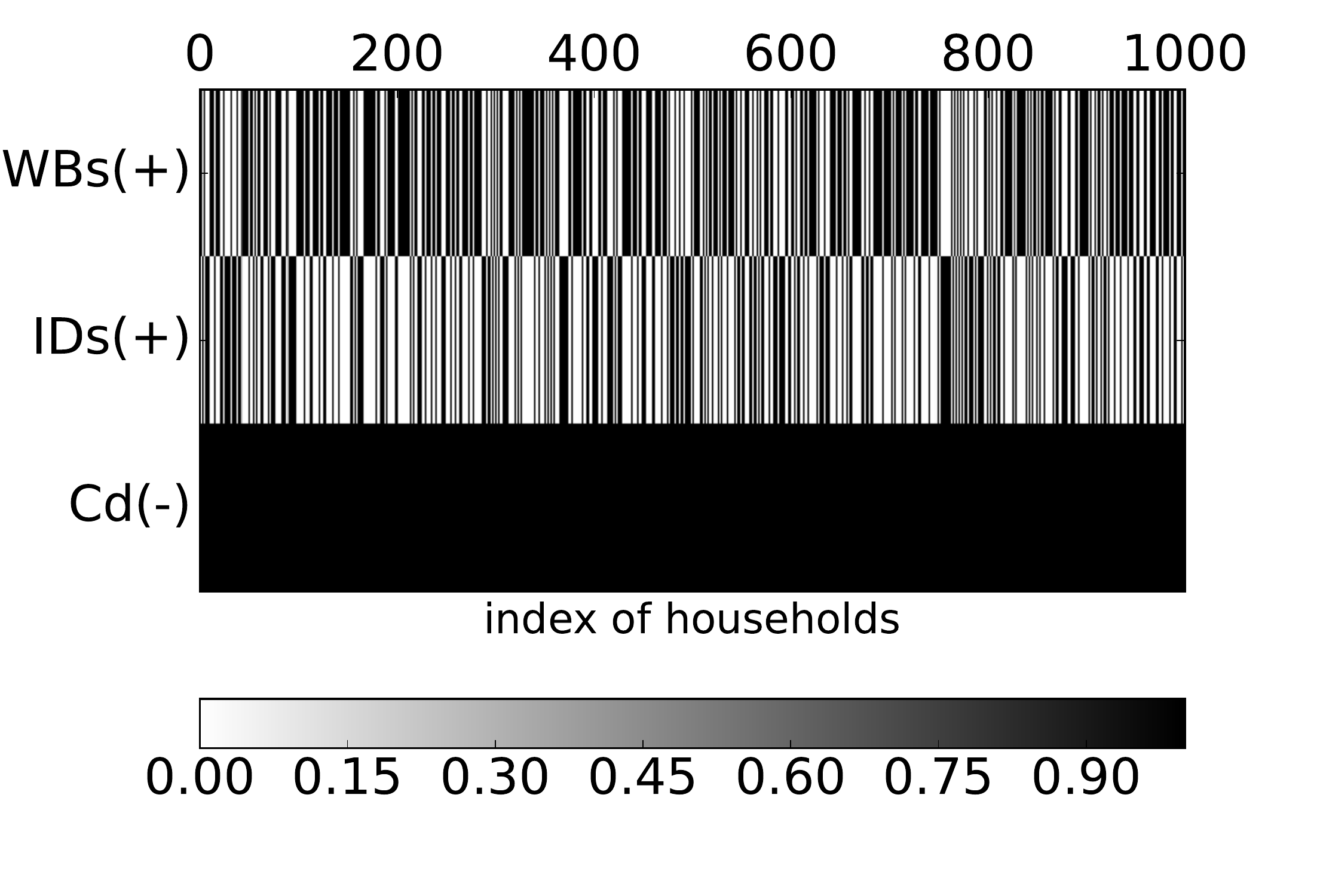}}
\subfigure[~]{
\includegraphics[width=7cm]{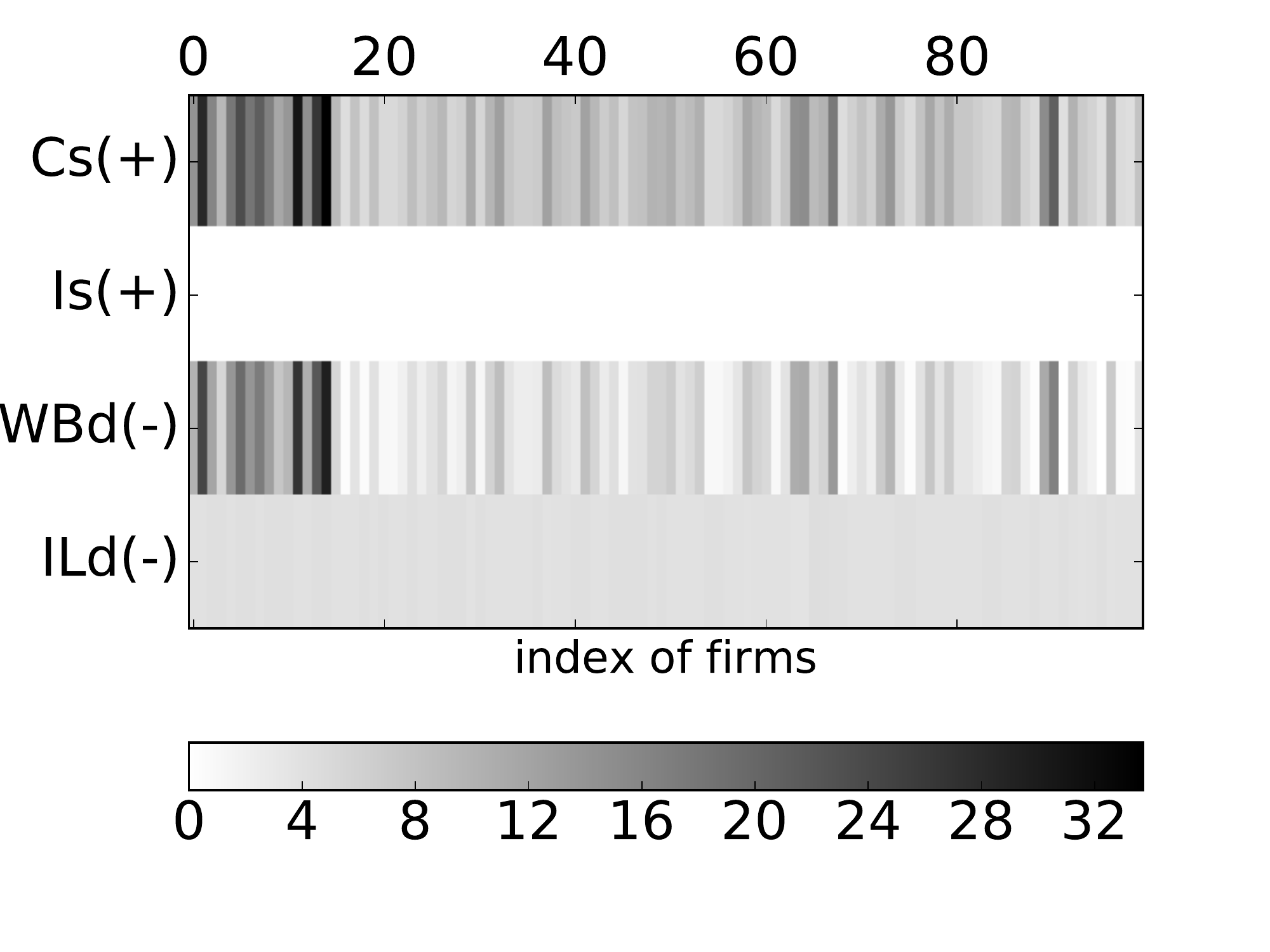}}
\caption{Budget of (a) households; (b) firms. $nb=3, ~nf=100, ~nh=1000$}
\label{fig.budget}
\end{figure}

Fig. \ref{fig.budget}(a) represents the budget of all households,
with a fixed consumption demand $\alpha_0$ imposed by $A_1$ and $b_1$, as a
minimalist way to ensure nonhomogeneity.
The population of agents can be clustered in two groups: their
consumption is either financed by wage bills or by interests on 
deposits, and this two values are negatively correlated. 
Furthermore the values taken by WBs and IDs are highly clustered 

Fig. \ref{fig.budget}(b) represents the budget of firms.
Income is mainly composed of consumption supply to households,
which is much larger than investment supply to firms.
Expense includes wage bills and interests on loans, the 
former dominating the latter.
We note also that $WBd$ and $Cs$ are positively correlated.
There is a sectorial dependence of $Cs$ that can be observed 
more in detail in Fig.\ref{fig.scatter.Cs.outdegree}

\subsection{Relationship between topological properties, stocks and flows 
in the block model}
\label{sub.valid.flow}


In the present section $n_{iter}$ networks are sampled from the
ensemble defined by the random fitness model. For each of them,
the approximate NNLS solution to eq.(\ref{eq.nnls.alpha0}) is computed. 
Then the values of $\xi$ corresponding to various economic transactions
are compared to the topological features of the related subnetworks
established in sec. \ref{sub.valid.topo}.

\begin{figure}[htbp]
\centering
\includegraphics[width=14cm]{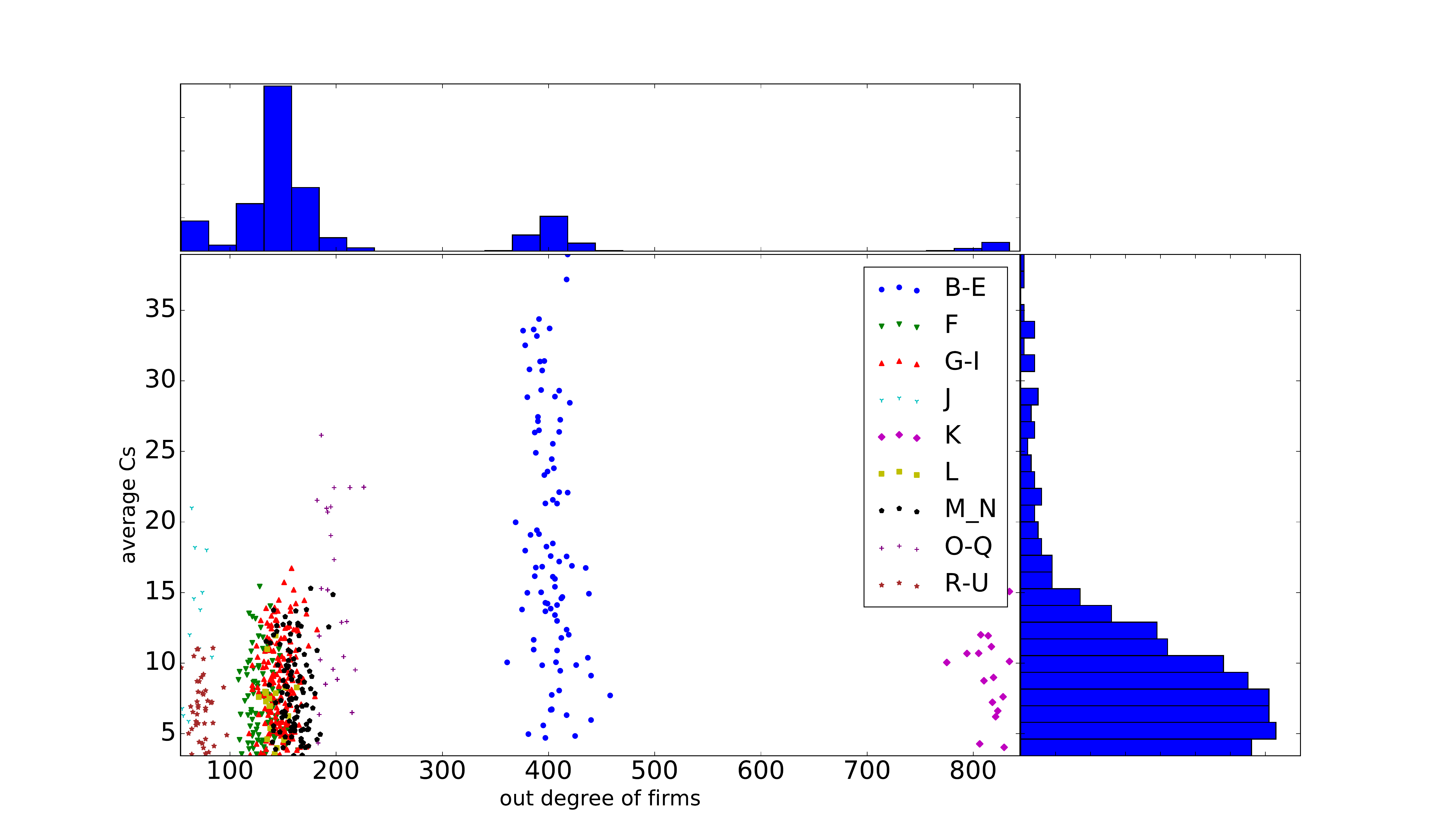}
\caption{Consumption supply $C_s$ from firms to households
and out-degree of the consumption network.  
(center) Each dot corresponds to a specific firm and represents its
out-degree, $C_s$ computed using NNLS, and its sector;
(top) marginal histogram of out-degree;
(right) marginal histogram of $C_s$. $n_b=3, n_f=50, n_h=300, n_{iter}=10$}
\label{fig.scatter.Cs.outdegree}
\end{figure}

The consumption $ C_s$ supplied to households by firms is shown in 
Fig. \ref{fig.scatter.Cs.outdegree} with respect to the out-degree of
firms, that is the number of their customers among households.
The distribution of sectors along the $x$-axis is well clustered and 
reflects the distribution of the connection probability, seen in 
Fig. \ref{fig.topological.properties.cons.hh}, for example the high
probability of connection of sector K.
However, this clustering effect is attenuated
by the uniform random fitness model in comparison to the block
model defined in sec. \ref{sub.application.network.bmw}.
Given the sector, it seems that $C_s$ is uniformly distributed
on some interval $[u,v]$. Comparing sectors B-E and K, $v$ 
does not appear to be a linear function of the sector's
connection probability $p_{s_i}$.

\begin{figure}[htbp]
\centering
\includegraphics[width=14cm]{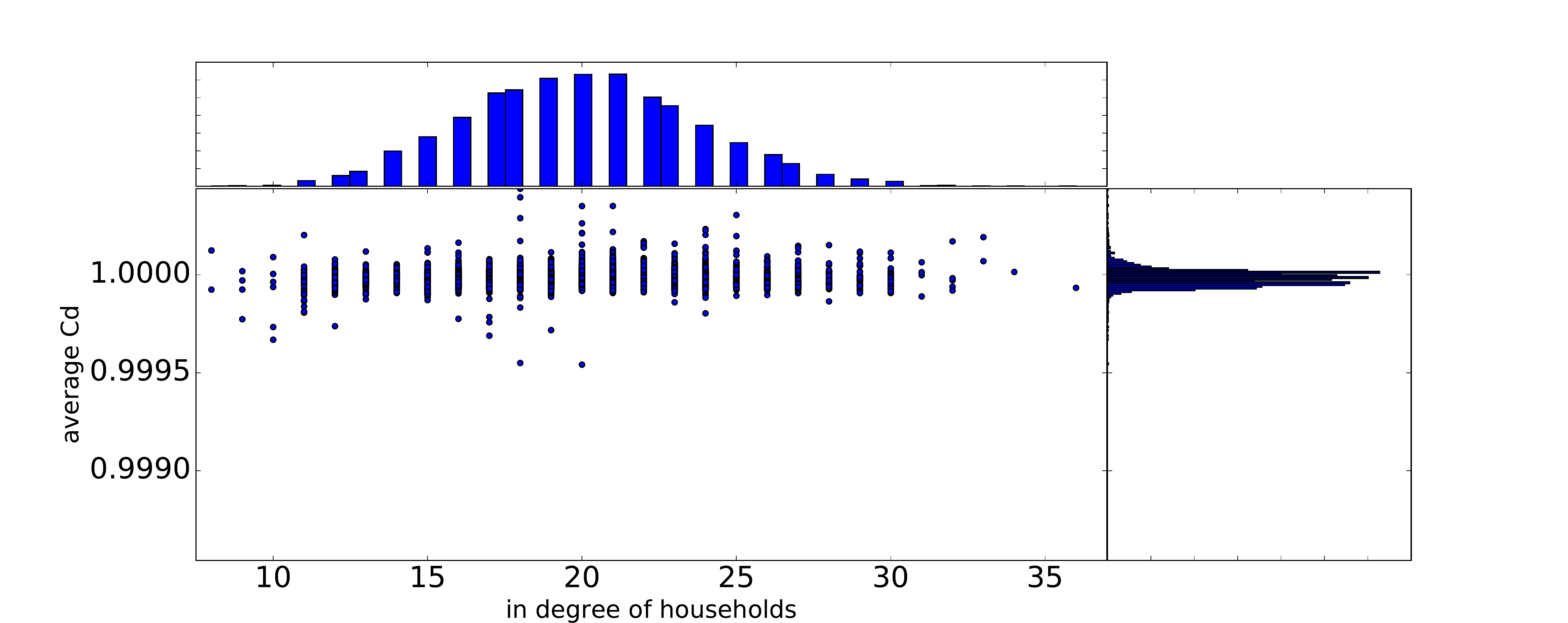}
\caption{Consumption demand $C_d$ by households and
in-degree of the consumption network.  
(center) Each dot corresponds to a specific household and represents
its average $C_d$ computed using NNLS and its in-degree ;
(top) marginal histogram of in-degree ;
(right) marginal histogram of $C_d$.  $n_b=3, n_f=100, n_h=1000, n_{iter}=10$}
\label{fig.scatter.Cd.indegree}
\end{figure}

The average consumption $C_d$ demanded by households is
represented in Fig. \ref{fig.scatter.Cd.indegree} with respect to
the in-degree of households, that is the number of firms a given 
household is buying from. 
It can be verified that households form a homogeneous group in the
block model and the random fitness models. Furthermore they
can't be associated to an industrial sector in a one-to-one map.
The clustering around the value of $C_d$ imposed by the rhs $b_1$ 
in eq.(\ref{eq.nnls.alpha0}) is clearly observed.

In sec. \ref{sec.budget}, it was remarked that the investment 
$I_s$ between firms had a residual level compared to other quantities.
As a first explanation, it can be noticed that the number of inter-firm
links is very low compared to firm-household links.
This problem is likely to disappear when adding capital depreciation 
equations that constrain the investment level, as in the BMW model.

\section{Comparison to other methods}
\label{sec:comparison}

In sec. \ref{sec.marginal.pdf}, numerical NNLS solutions to the
problem in eq.(\ref{eq.nnls.alpha0}) were computed, as the network topology
was sampled from a probabilistic FiCM model. In this section, our
approach, noted FiCM+NNLS, is compared to other existing works. 

Firstly, the ``degree-corrected gravity model'' (FiCM+dCGM) is a two-step
method that builds on an FiCM model \cite{cimini_systemic_2015, squartini_enhanced_2017}, 
inspired by the gravity model, and corrected in order to set a given level
of sparsity, and to apply a constraint on the degree sequence.
The value of weights placed on edges is:
\begin{eqnarray} 
w_{ij}&=&\frac{a_{ij}}{W}(z^{-1}+ s_i s_j)
\end{eqnarray} 
where $W$ is a normalization factor,  $s_i=\sum_j w_{ij}$ and $s_j=\sum_i w_{ij}$
are the nodes' strengths. For the sake of comparison, $W_i, ~W_j$ and $W$ are computed below using
the NNLS solution.

Secondly, FiCM+NNLS will be compared to the Bayesian method 
in \cite{waldrip_comparison_2017} that results in eq.(\ref{eq.solution.bayes}).
All methods give $w_{ij}$ conditioned on the knowledge of the
connection probabilities $a_{ij}$ of the subgraphs.
Even though FiCM+dcGM and FiCM+NNLS build probabilistic models of topologies,
expressed by $p_{ij}$, they compute connection weights 
conditionaly on a particular realization of $a_{ij}$, as explained above.
Similarly, the Bayesian method in \cite{waldrip_comparison_2017} 
computes the mean $\langle \xi \rangle$ using a closed-form expression
that needs $a_{ij}$ to be known.
We will be interested by several features of the obtained solutions: 
\begin{itemize}
\item do they return negative money flows (that are not compatible with the 
desired behavior) ? This is quantified by the percentage of negative
coefficients over non-zero coefficients in $\xi$.
\item to what extend do they respect the linear system of equations in eq.(\ref{eq.CSP}) ?
This is quantified by a relative error defined by $100\times\frac{\|A_1\xi-b_1 \|_1}{\|\xi\|_1}$.
\end{itemize}
Tab. \ref{tab.comparison} summarizes a computational comparison based
on repreating $n=100$ times each method, and estimating the
sample value of the error indicators.
As expected, since FiCM+dcGM is not designed to impose the linear 
constraint $A_1 \xi=b_1$, the relative error rate obtained is much higher
than the value obtained with the other methods. FiCM+NNLS and the Bayesian
algorithm yield a satisfactory relative residual error below $1 \%$.
Then, the rate of negative coefficients is compared, and we find a value
as high as 21\% for the Bayesian method. This was anticipated in 
sec. \ref{sec.marginal.pdf}, but is not acceptable in our context.
Potential solutions to this issue will be discussed in sec. \ref{sec.discussion}.

Some other features can be remarked: first the FiCM+dcGM and Bayesian method
yield a closed-form expression for $w_{ij}$ conditioned on $a_{ij}$,
unlike FiCM+NNLS. 
There are method, such as the ECM \cite{mastrandrea_enhanced_2014}, that
outputs both $P(a_{ij}=a)$ and $P(w_{ij}=w)$ in a closed-form way, but 
with the same drawback as FiCM+dcGM since there is no way to impose 
the condition $A_1 \xi=b_1$.

To conclude this section, FiCM+NNLS stands as the only method among
the considered ones that respects both non-negativity of coefficients and
the linear system of equations $A_1 \xi=b_1$.
In the following section, other possible improvements or research 
directions will be discussed.

\begin{table}[htbp]
\centering
\begin{tabular}{p{3.5cm}p{4cm}p{4cm}p{4cm}}  
\hline
Method & FiCM+dcGM  & FiCM+NNLS & Bayesian \\
\hline	
\hline
Input & Fitnesses $x_i,y_j$, number of links $L$, $W_i,W_j$ for each subnetwork. & Fitnesses $x_i,y_j$, number of links $L$ for each subnetwork. $A_1,b_1$ corresponding to $a_{ij}$ & Network topology $a_{ij}$.  $A_1,b_1$ corresponding to $a_{ij}$ \\
\hline
Probabilistic model of topology $p_{ij}$ & Yes & Yes & No \\
\hline
Closed-form expression of $w_{ij}$ & Yes, conditioned on $a_{ij}$. & No & Yes, conditioned on $a_{ij}$. \\
\hline
Relative error \% & $15.9$& $0.18$  & $0.15$  \\
\hline
Negative coefficients \% &  $0.$ & $0.$ & $21.1$ \\
\hline
Refs & \cite{squartini_enhanced_2017} & Present article & \cite{waldrip_comparison_2017} \\
\hline
Comments & High error since $A_1 \xi=b_1$ is not taken into account. & No pdf is available for $w_{ij}$.  & Outputs negative coefficients. Estimation can be augmented with observed values. \\
\hline
\end{tabular}
\caption{\label{tab.comparison} Comparison of ensemble reconstruction methods.
Relative error is defined by $100\times\frac{\|A_1\xi-b_1 \|_1}{\|\xi\|_1}$ and averaged over $n=100$ trials. 
Negative coefficients \% is the percentage of negative coefficients over non-zero coeffecients of $\xi$.
}
\end{table}	

\section{Discussion}
\label{sec.discussion}

The main idea developed in this article is to build a systemic macroeconomic
model able to reproduce topological features, with respect to a given 
theoretical behavior. This explains the choice of a minimalist set
of behavioral equations, that may be extended to include 
other features of SFC models. The interplay between stocks
(wealth of households $M$, stock of loans $L$, capital stock of firms $K$)
and flows was not examined here. It is known from empirical
studies that the network of interfirm sales should be disassortative
\cite[§2]{letizia_corporate_2018} but we could not observe realistic
values for $Is$ as explained in sec. \ref{sub.valid.flow}. 
Apart from linear models, nonlinearities may be introduced, following the
example given in engineering \cite{waldrip_maximum_2016}.

In sec. \ref{sub.application.network.bmw}, several topological models were 
proposed for each transaction subnetwork. It was noticed that some were not
consistent with empirical evidence. Among FiCM networks, the block model
for investment and consumption of firms was modified to account
for heterogeneity among sectors. A uniform random fitness model was proposed
and can be extended to power-laws, building on studies of firm size \cite{axtell_zipf_2001,gabaix_power_2009}.
The scope could be extended beyond national limits, following \cite{mizuno_structure_2016}.
The network of household consumption should incorporate more empirical
data and notably the different socioeconomic positions \cite{leo_correlations_2018}.
The BiRG networks may be replaced by appropriate FiCM networks. The network of
bank loans to firms should use central bank data \cite[3.2]{silva_bank_2017}.
The network of wages could benefit from aggregate constraints such that the
compensation of employees in national accounts \cite[5.1]{eurostat_:_statistisches_amt_der_europaischen_gemeinschaften_eurostat_2008}.
The investment network, which was proxied by firm consumption, could be
enhanced using a better data source. Agriculture, which was not excluded 
in the network model for convenience, should be included.

The experimental analysis of the reconstructed networks' properties in
sec. \ref{sub.valid.topo} and \ref{sub.valid.flow}
consists in a comparison to basic stylized facts in the empirical litterature.
This validation, while necessary is not sufficient and should be extended
in two possible directions: either with a comparison to detailed micro-data
from empirical studies, of from ABM simulation where the real topology
is known. 

The two-step method devised here first needs to reconstruct the topology from
economic data. Then, weights are approximated using an 
numerical method. Those weights are then expected to reflect the economic
data used in the first step, which can be validated experimentally (but is
not guaranteed by construction). 
A single-step method similar to \cite{mastrandrea_enhanced_2014}
and able to cope with the constraint in eq. (\ref{eq.syslin}), could
provide such a guarantee. Some recent works \cite{polpo_maximum_2018}
in that direction question the relevant form for the entropy. 
Also, maximum entropy methods with inequality constraint should be
taken into account.

The role of time may be questionned, as it is expected to play no role in
this article, on the opposite of ABM that can deal with growth and transient
phenomena. Small fluctuations in the vicinity of the data points used 
to learn the subnetworks can be discussed, but it is unlikely that radical
change in the network structure can be modelled. Stability issues 
close to the steady-state can be examined using the theory of dynamics
on networks \cite[18.2]{newman_networks:_2010},\cite{anand_stability_2009}.
More importantly, it must be stressed that the concept of steady-state 
\cite{daly_uneconomic_2014} has been debated during several decades in 
the field of ecological economics. The benefits of using our approach
in that context will be examined.

A detailed analysis of our results in the light of economic knowledge is
needed. For example the influence of the macroeconomic parameters
on the network structure may be compared to what is expected by existing
economic theories, and empirical observations.


	
	
		
\section{Conclusion}
\label{sec.conclusion}

In this article we propose an intermediate model between ABM and
complex networks, able to reflect topological features, 
heterogeneity and interaction, with theoretical properties that are
easier to establish than in the ABM case. This comes at the cost 
of losing time-dependence.

Data-driven economic network reconstruction methods
do not include so far constraints stemming from macroeconomic models. 
In this article we propose to introduce such a constraint, that induces
a specific distribution for the weights of the network.
To do so, we use a two-step method that requires first to estimate the
topology of each subnetwork taken independently, then to 
estimate network weights. The first step can be skipped if a detailed
empirical description of the network topology is known.

Building on the fitness-induced configuration model we defined several
connection probabilities to model the topology of subnetworks representing
various economic transactions. These models respect the empirical link density
found in empirical studies and were fit to national accounts empirical data.

In the future we will extend the methods developed here in several 
direction: first, instead of a two-step approach, we plan to get the
parameters of the full network in just one step. Furthermore, 
reconstructed networks will be compared to a ground truth obtained
from empirical or simulated data. More detailed economic models,
possibly non-linear, can be studied.

Lastly we stress that the results obtained can be used by 
practitioners in the ABM or SFC communities that are interested
in network and distributional phenomena in the steady state.

\appendix

\section{Notations}
\label{appendix.notations.syslin}

In this section the notations of sec. \ref{sec.sfc.model} are explicited.

\begin{equation}   
\begin{split} 	
WBd & =
\begin{bmatrix}
WBd_{1,1} & \ldots &WBd_{n_h,1} & WBd_{1,2} & \ldots &WBd_{n_h,2} &  \ldots& WBd_{1,n_f} & \ldots &WBd_{n_h,n_f} 
\end{bmatrix} \\
IDd &=
\begin{bmatrix}
IDd_{1,1} & \ldots &IDd_{n_h,1} & IDd_{1,2} & \ldots &IDd_{n_h,2} &  \ldots& IDd_{1,n_b} & \ldots &IDd_{n_h,n_b} 
\end{bmatrix} \\
ILd&=
\begin{bmatrix}
ILd_{1,1} & \ldots &ILd_{n_h,1} & ILd_{1,2} & \ldots &ILd_{n_h,2} &  \ldots& ILd_{1,n_b} & \ldots &ILd_{n_h,n_b} 
\end{bmatrix}\\
Id&=
\begin{bmatrix}
Id_{1,1} & \ldots &Id_{n_f,1} & Id_{1,2} & \ldots &Id_{n_f,2} &  \ldots& Id_{1,n_f} & \ldots &Id_{n_f,n_f} 
\end{bmatrix} 
\end{split}  
\end{equation}
where $WBd_{n_h,n_f}$ is the wage obtained by household $n_h$ from
firm $n_f$, $IDd_{n_h,n_b}$ is the amount of interest obtained by
household $n_h$ from bank $n_b$, $ILd_{n_h,n_b}$ is the amount of
interest paid by firm $n_h$ to bank $n_b$, 
$Id_{n_{f,1},n_{f,2}}$ is the investment paid by firm $n_{f,1}$
to firm $n_{f,2}$.


\section{Eurostat national accounts databases}
\label{appendix.supply.use}


All tables in this section are built by national accountants for a given country,
and a given year, but these mentions are dropped for clarity reasons.

The production matrix forms a part of the supply table. For each category of products in
the rows, it displays the value of the production, grouped by industry type in the columns.
To simplify, only the production matrix is shown in Tab. \ref{tab.supply} while the 
others components of the supply table are dropped (no imports, trade and transport margins,
taxes less subsidies on products) \cite[Tab. 4.3, §4.1]{eurostat_:_statistisches_amt_der_europaischen_gemeinschaften_eurostat_2008}.
The value of product $p \in [1,P]$ produced by sector $s \in S$ will be noted $sup[p,s]$.

\begin{table}[htbp]
\centering
	\begin{tabular}{ll|lll|l} %
	\hline
	 &  &  \multicolumn{3}{c}{Output of industries}  &  \\
	 \hline
	 & Industries  & Agriculture & \ldots & Other services & Total domestic output \\
	 Products &  & 1 & \ldots & $n$ &   \\
	\hline
	\hline
	Products of agriculture &  &  & \ldots & &   \\
	\vdots &  &  & \ldots & & $\Sigma$   \\
	Products of other services &  &  & \ldots & &   \\
	\hline
	Total &  &  & $\Sigma$  & &   \\
	\hline
	\end{tabular}
\caption{\label{tab.supply} Production matrix, that constitutes the first quadrant of the supply table.}
\end{table} 

As explained in \cite[§5.1]{eurostat_:_statistisches_amt_der_europaischen_gemeinschaften_eurostat_2008}
``a use table shows the use of goods and services by product
and by type of use for intermediate consumption by industry,
final consumption expenditure, gross capital formation or exports''.
Only the quadrant named ``Final uses'' will be used here, and
more particularly two columns, ``Final consumption expenditure by households'' 
and ``Gross fixed capital formation''. This is summarized in Tab.\ref{tab.use}.
Since our model does not involve intermediate consumption, the corresponding 
part of the use table is not exploited here.
The value of product $p \in [1,P]$ consumed by households as a final use is
noted $use^{fin}[p]$.

The column ``Gross fixed capital formation'' can in fact be disaggregated by 
investing industry \cite[Fig 5.1 p.125]{eurostat_:_statistisches_amt_der_europaischen_gemeinschaften_eurostat_2008},
and is called the Investment matrix.
The amount of fixed capital of product $p \in [1,P]$ formed by the industrial
sector $s \in S$ is written $use^{cap}[p,s]$.

\begin{table}[htbp]
\centering
	\begin{tabular}{ll|p{4cm}p{4cm}l|l} %
	\hline
	 &  &  \multicolumn{3}{c}{Final uses}  &  \\
	 \hline
	  &   & Final consumption expenditure by households & Gross fixed capital formation & \ldots & Total \\	  
	 Products &  &  &  & &   \\
	\hline
	\hline
	Products of agriculture &  &  & \ldots & &   \\
	\vdots &  &  & \ldots & &  $\Sigma$  \\
	Products of other services &  &  & \ldots & &   \\
	\hline
	Total &  &  & $\Sigma$  & &   \\
	\hline
	\end{tabular}
\caption{\label{tab.use} Final uses matrix, that constitutes the second quadrant of the use table.}
\end{table}

The rows of the industry-by-industry input-output table in Tab.\ref{tab.io}
explain how the production of a given sector is sent to other sectors.
The columns show the different inputs of a given sector.

\begin{table}
\centering
\begin{tabular}{llll|l}
\hline
& Agri. & Industry & Services  & Uses \\
\hline
Agriculture	& 10& 34  &10 &\\
Industry	&20 & 152& 40 & $\Sigma$\\
Services  &10 & 72& 20 &\\
\hline
Output  & & $\Sigma$ &  \\
\end{tabular}
\caption{\label{tab.io} Input-output matrix, industry-by-industry.}
\end{table}

Tab. \ref{tab.datasets} summarizes the data sources used in this article.

\begin{table}[htbp]
\centering
	\begin{tabular}{p{3.5cm}p{3.5cm}p{3.5cm}p{3.5cm}} %
	\hline
	Dataset label & Dataset definition & Input for transaction networks & Corresponding variable or fitness \\	
	\hline	
	\hline
	bd\_9ac\_l\_form\_r2  & business demography by legal form &Investment, consumption, wage& $nf,nh$ by sector \\		
	\hline
	naio\_10\_cp15, naio\_10\_cp16, naio\_10\_cp1750 & supply,use, input-output tables & Investment & $d_{s_i s_j}$ \\
	\hline	
	 &  & Consumption  & $x_{s_i}$ \\
	\end{tabular}
\caption{\label{tab.datasets} Eurostat datasets used to parametrize random networks}
\end{table}

 \begin{table}[htbp]
\centering
\begin{tabular}{ll}
\hline
Label&Definition \\
\hline
A& agriculture, forestry and fishing \\
B& mining and quarrying \\
C& manufacturing \\
D& electricity, gas, steam and air conditioning supply \\
E& water supply; sewerage, waste management and remediation activities \\
F& construction \\
G& wholesale and retail trade; repair of motor vehicles and motorcycles \\
H& transportation and storage \\
I& accommodation and food service activities \\
J& information and communication \\
K& financial and insurance activities \\
L& real estate activities \\
M& professional, scientific and technical activities \\
N& administrative and support service activities \\
O& public administration and defence; compulsory social security \\
P& education \\
Q& human health and social work activities \\
R& arts, entertainment and recreation \\
S& other service activities \\
T& activities of households as employers or for own use \\
U& activities of extraterritorial organisations and bodies 
\end{tabular}
\caption{Definition of sectors}
\label{tab.sector.labels}
\end{table}  
  
\section{Acknowledgements}
We thank two anonymous reviewers for their comments that helped 
improving the manuscript.
Open-source software were used to perform this research: Python, 
Scipy, \LaTeX, Matplotlib.







\end{document}